\documentclass[11pt,a4paper]{article}
\usepackage{jcappub}
\usepackage{lmodern}
\usepackage{graphicx,color}
\graphicspath{{figures/}}
\usepackage{comment}
\usepackage{cancel}

\usepackage[T1]{fontenc}
\usepackage[utf8]{inputenc}
\usepackage{color}
\usepackage{amsmath}
\usepackage{amssymb}
\usepackage{esint}
\usepackage{appendix}
\usepackage{bm}
\usepackage{url}
\usepackage{hyperref}
\usepackage[normalem]{ulem}

\makeatletter

\definecolor{somegreen}{cmyk}{0,0.49,0.98,0.09}
\definecolor{red}{rgb}{1,0,0}
\definecolor{magenta}{cmyk}{0,1,0,0}
\definecolor{lavender}{cmyk}{0.16,0.67,0,0.57}
\definecolor{darkgreen}{rgb}{0,0.65,0.05}
\definecolor{antiquefuchsia}{rgb}{0.33, 0.1, 0.89}

\definecolor{airforceblue}{rgb}{0.36, 0.54, 0.66}

\def\beqra{\begin{eqnarray}}
\def\eeqra{\end{eqnarray}}
\def\beq{\begin{equation}}
\def\eeq{\end{equation}}

\def\vp{\bar{\varphi}}

\def\vp{\varphi}

\def\bx{{\bf{x}}}
\def\bz{{\bf{z}}}
\def\bk{{\bf{k}}}
\def\bp{{\bf{p}}}
\def\bq{{\bf{q}}}

\def\bv{{\bf{v}}}

\def\bV0{{\bf{V_0}}}
\def\re#1{(\ref{#1})}

\def\bx{{\bf{x}}}

\def\bs{{\bf{s}}}
\def\bk{{\bf{k}}}
\def\bp{{\bf{p}}}
\def\bq{{\bf{q}}}

\def\bv{{\bf{v}}}
\def\bV{{\bf{V}}}

\def\bz{{\bf{z}}}

\def\phil{\phi^{(1)}}

\def\vp{\varphi}
\def\vpl{\varphi^{(1)}}

\let\jnl@style=\rm
\def\ref@jnl#1{{\jnl@style#1}}

\def\aj{\ref@jnl{AJ}}                   
\def\actaa{\ref@jnl{Acta Astron.}}      
\def\araa{\ref@jnl{ARA\&A}}             
\def\apj{\ref@jnl{ApJ}}                 
\def\apjl{\ref@jnl{ApJ}}                
\def\apjs{\ref@jnl{ApJS}}               
\def\ao{\ref@jnl{Appl.~Opt.}}           
\def\apss{\ref@jnl{Ap\&SS}}             
\def\aap{\ref@jnl{A\&A}}                
\def\aapr{\ref@jnl{A\&A~Rev.}}          
\def\aaps{\ref@jnl{A\&AS}}              
\def\azh{\ref@jnl{AZh}}                 
\def\baas{\ref@jnl{BAAS}}               
\def\bac{\ref@jnl{Bull. astr. Inst. Czechosl.}}
\def\caa{\ref@jnl{Chinese Astron. Astrophys.}}
\def\cjaa{\ref@jnl{Chinese J. Astron. Astrophys.}}
\def\icarus{\ref@jnl{Icarus}}           
\def\jcap{\ref@jnl{J. Cosmology Astropart. Phys.}}
\def\jrasc{\ref@jnl{JRASC}}             
\def\memras{\ref@jnl{MmRAS}}            
\def\mnras{\ref@jnl{MNRAS}}             
\def\na{\ref@jnl{New A}}                
\def\nar{\ref@jnl{New A Rev.}}          
\def\pra{\ref@jnl{Phys.~Rev.~A}}        
\def\prb{\ref@jnl{Phys.~Rev.~B}}        
\def\prc{\ref@jnl{Phys.~Rev.~C}}        
\def\prd{\ref@jnl{Phys.~Rev.~D}}        
\def\pre{\ref@jnl{Phys.~Rev.~E}}        
\def\prl{\ref@jnl{Phys.~Rev.~Lett.}}    
\def\pasa{\ref@jnl{PASA}}               
\def\pasp{\ref@jnl{PASP}}               
\def\pasj{\ref@jnl{PASJ}}               
\def\rmxaa{\ref@jnl{Rev. Mexicana Astron. Astrofis.}}%
\def\qjras{\ref@jnl{QJRAS}}             
\def\skytel{\ref@jnl{S\&T}}             
\def\solphys{\ref@jnl{Sol.~Phys.}}      
\def\sovast{\ref@jnl{Soviet~Ast.}}      
\def\ssr{\ref@jnl{Space~Sci.~Rev.}}     
\def\zap{\ref@jnl{ZAp}}                 
\def\nat{\ref@jnl{Nature}}              
\def\iaucirc{\ref@jnl{IAU~Circ.}}       
\def\aplett{\ref@jnl{Astrophys.~Lett.}} 
\def\apspr{\ref@jnl{Astrophys.~Space~Phys.~Res.}}
\def\bain{\ref@jnl{Bull.~Astron.~Inst.~Netherlands}}
\def\fcp{\ref@jnl{Fund.~Cosmic~Phys.}}  
\def\gca{\ref@jnl{Geochim.~Cosmochim.~Acta}}   
\def\grl{\ref@jnl{Geophys.~Res.~Lett.}} 
\def\jcp{\ref@jnl{J.~Chem.~Phys.}}      
\def\jgr{\ref@jnl{J.~Geophys.~Res.}}    
\def\jqsrt{\ref@jnl{J.~Quant.~Spec.~Radiat.~Transf.}}
\def\memsai{\ref@jnl{Mem.~Soc.~Astron.~Italiana}}
\def\nphysa{\ref@jnl{Nucl.~Phys.~A}}   
\def\physrep{\ref@jnl{Phys.~Rep.}}   
\def\physscr{\ref@jnl{Phys.~Scr}}   
\def\planss{\ref@jnl{Planet.~Space~Sci.}}   
\def\procspie{\ref@jnl{Proc.~SPIE}}   

\makeatother

\title{Bootstrapping Lagrangian Perturbation Theory for the Large Scale Structure
}
\author[a]{Marco Marinucci,}
\author[b]{Kevin Pardede,}
\author[b,c]{Massimo Pietroni}
\affiliation[a]{Dipartimento di Fisica e Astronomia “G. Galilei”, Università degli Studi di Padova and INFN, Sezione di Padova, via Marzolo 8, I-35131, Padova, Italy}
\affiliation[b]{INFN, Gruppo Collegato di Parma, Parco Area delle Scienze 7/A, I-43124, Parma, Italy}
\affiliation[c]{Dipartimento di Scienze Matematiche, Fisiche e Informatiche, Universit\`a di Parma, Parco Area delle Scienze 7/A, I-43124, Parma, Italy}

\abstract{We develop a model-independent approach to lagrangian perturbation theory for the large scale structure of the universe. We focus on the displacement field for dark matter particles, and derive its most general structure without assuming a specific form for the equations of motion, but implementing a set of general requirements based on symmetry principles and consistency with the perturbative approach. We present explicit results up to sixth order, and provide an algorithmic procedure for arbitrarily higher orders. The resulting displacement field is expressed as an expansion in operators built up from the linear density field, with time-dependent coefficients that can be obtained, in a specific model, by solving ordinary differential equations. The derived structure is general enough to cover a wide spectrum of models beyond $\Lambda$CDM, including modified gravity scenarios of the Hordenski type and models with multiple dark matter species.
This work is a first step towards a complete model-independent lagrangian forward model, to be employed in cosmological analyses with power spectrum and bispectrum, other summary statistics, and field-level inference.} 
\emailAdd{marco.marinucci@unipd.it}
\emailAdd{kevinfranklysamuel.pardede@unipr.it}
\emailAdd{massimo.pietroni@unipr.it}
\begin{document}

 \maketitle
\section{Introduction}

The ongoing stage-IV galaxy surveys such as \textsc{Euclid} \cite{2009arXiv0912.0914L} and \textsc{DESI} \cite{DESI:2016fyo} are expected to provide a significant amount of cosmological data. These data will be crucial for addressing important questions about our universe, such as the origin of cosmic acceleration, the behaviour of gravity at cosmological scales, the nature of massive neutrinos and dark matter particles, and the statistical properties of the universe's initial conditions.

To extract cosmological information from  galaxy surveys it is essential to model the non-linearity of galaxy clustering. While N-body simulations provide precise modeling of galaxy clustering at the nonlinear scale, they are generally computationally expensive. Moreover, to simulate the statistics of galaxies (the objects we observe), we need hydrodynamical codes, which are less reliable due to the highly complex physics of galaxy formation. However, it is worth noting that there has been significant progress on emulators trained to produce various observables predicted by N-body simulations reliably and quickly (see for instance \cite{Lawrence:2009uk, Nishimichi:2018etk, Liu:2017now, Euclid:2018mlb, DeRose:2018xdj, Angulo:2020vky}).

A complementary approach is to provide an analytical treatment to the nonlinearities. In recent years, several different analytic frameworks have been put forward \cite{Bernardeau:2001qr, RPTa, Matarrese:2007wc, Taruya2007, Pietroni:2008jx, Bernardeau:2008fa, Baumann:2010tm, Pietroni:2011iz, Carrasco:2012cv, Manzotti:2014loa, Senatore:2014via, Baldauf:2015xfa, Blas:2015qsi, Blas:2016sfa, Peloso:2016qdr, Noda:2017tfh, Ivanov:2018gjr,Moutarde:1991evx, Buchert:1992ya, Buchert:1993ud, Bouchet:1994xp, Catelan:1994ze, Porto:2013qua, Vlah:2015sea}
The basic ingredient of these frameworks is perturbation theory (PT) (see \cite{Bernardeau:2001qr, Bernardeau:2013oda, Desjacques:2016bnm} for review). In a first step, the nonlinear dark matter field is expressed as an expansion of the initial density field. 
Then, since the PT expansion fails in accounting for the effect of short (UV) scales on the intermediate ones relevant to galaxy surveys,  the theory is corrected by adding ``UV counterterms'' \cite{Baumann:2010tm, Pietroni:2011iz, Carrasco:2012cv}. These are  organized in a derivative expansion on long wavelength fields with phenomenological coefficients encoding the non-perturbative UV effects. Finally, the relation between the observed galaxy fields and the underlying dark matter one is modelled through the bias expansion, which is also an expansion in derivatives of the gravitational potential and the velocity field with coefficients parameterizing the  physics of galaxy formation (for a review, see \cite{Desjacques:2016bnm}). 

The purpose of this paper is to formulate the PT  expansion in a model-independent way, that is, without assuming a specific form of the time evolution equations, but only a set of underlying symmetries, supplemented by a few other basic requirements. In standard PT approaches, 
the nonlinear fields are obtained, in a given model, by an iterative solution of the nonlinear evolution equations in that model. The result is expressed, at $n$-th order in PT, as a convolution (in momentum space) or the product (in configuration space) of $n$ initial fields. 
As is well known, at first order the solution of the linearized equations gives the linear growth factor, $D$, and its logarithmic derivative, the growth function $f$, which in General Relativity (GR) can be approximated as
\beq
f =\frac{ d \log D}{d \log a}\simeq \Omega_m(a)^\gamma\,,
\label{eq:gammapar}
\eeq
where $a$ is the scale factor and $\Omega_m(a)$  the time-dependent matter density, normalized to the critical density. The exponent $\gamma$ can be approximated as $\gamma \simeq 0.55$ for $\Lambda$CDM, while it receives corrections in models in which the Dark Energy is not a cosmological constant \cite{Linder:2005in}. Measurements of $f(a)$, for instance from the Kaiser effect \cite{Kaiser:1987qv}, can then provide model-independent tests of $\Lambda$CDM on cosmological scales (see, for instance, \cite{Simpson:2009zj, Beutler:2012px, Song:2015oza, BOSS:2016wmc}). 
Recently, many models beyond $\Lambda$CDM have been tested on galaxy clustering data from the BOSS survey~\cite{BOSS:2015npt}. The analyses on modified gravity~\cite{Piga:2022mge} (see also~\cite{Bose:2018orj}), interacting~\cite{Tsedrik:2022cri} and evolving dark energy~\cite{DAmico:2020kxu, DAmico:2020tty, Ivanov:2020ril, Chudaykin:2020ghx, Carrilho:2022mon} have  established the potential of large scale structure (LSS) in constraining fundamental physics. Intriguingly, the latest results from the BAO analysis of DESI data~\cite{DESI:2024mwx}, combined with cosmic microwave background and supernovae data, 
hint at a possible deviation from a cosmological constant in favour of an evolving dark energy, suggesting that new physics might be right behind the corner.

In view of the precision attainable from data by future surveys, it is therefore timely to extend these tests from the linear to the nonlinear sector, deriving constraints on the coefficients appearing in the most general PT expansion compatible with the assumed symmetries. 
Such a program was started in \cite{DAmico:2021rdb}, where it was dubbed `Large Scale Structure Bootstrap', in the same spirit of  `bootstrap' approaches popular nowadays in other fields, which aim at exploiting all possible constraints coming from symmetries, without relying on the explicit form of the equations of motion.  In \cite{DAmico:2021rdb}, the most general expressions, up to third order, for the PT kernels for dark matter density and velocity in momentum space was derived, after imposing  constraints coming from rotational invariance, Extended Galilean Invariance (EGI) \cite{Scoccimarro:1995if, Jain:1995kx, Peloso:2013zw, Kehagias:2013yd, Peloso:2013spa}, mass and momentum conservation. Moreover, it was shown that lifting the requirements of mass and momentum conservation, the kernels for biased tracers, equivalent to those of the perturbative bias expansions \cite{Desjacques:2016bnm}, were obtained. Previous work along similar lines was presented in \cite{Fujita:2020xtd}. The first Fisher forecast on the `bootstrap' parameters were performed in \cite{Amendola:2023awr,Amendola:2024gkz}, where it was shown that, combining power spectrum and bispectrum, an \textsc{Euclid}-like survey can   test   the nonlinear sector of GR at the 10\% level. 

The results of ref.~\cite{DAmico:2021rdb}, are obtained in the eulerian formulation of PT,  which computes the evolution of the density and velocity fields at any point, $\bx$, solving a system of continuity, Euler, and Poisson equations. As is well known, this formulation  is particularly advantageous for the computation of statistical correlators such as the power spectrum (PS) and the bispectrum (BS). 
In this paper, on the other hand, we adopt the equivalent lagrangian formulation of PT (LPT) \cite{Moutarde:1991evx, Buchert:1992ya, Buchert:1993ud, Bouchet:1994xp, Catelan:1994ze},  where the motion of fluid elements is followed in time, and is described by a {\it displacement  field} defined in lagrangian space, the space of the initial positions of the particles.
Recursive relations for the displacement field have been derived based on the equations of motion for the Einstein-de Sitter (EdS) or $\Lambda$CDM  models in  Refs.~\cite{2012JCAP...12..004R, Rampf:2015mza, Zheligovsky:2013eca, Matsubara:2015ipa, Baldauf:2015tla}. The results of this work, on the other hand, are based on symmetries rather than on the specific form of the equations of motion, and therefore extend previous results to any model sharing the same symmetries as EdS/$\Lambda$CDM.

As we will see, the lagrangian bootstrap features some relevant advantages with respect to the eulerian one. From the practical point of view, the implementation of the constraints, notably those coming from EGI, is much more straightforward, which allows us to derive an algorithmic procedure to derive the displacement field, in principle, to any PT order. Moreover, the lagrangian viewpoint is more natural, if one is interested in modeling the nonlinear fields instead of computing their correlators. This paper is therefore to be seen as a first step towards a model-independent implementation of cosmological inference  at field level \cite{Baldauf:2015tla,Taruya:2018jtk, Schmittfull:2018yuk, Schmidt:2018bkr, Elsner:2019rql, Schmidt:2020ovm, Schmittfull:2020trd, Taruya:2021ftd, Baumann:2021ykm, Andrews:2022nvv, Obuljen:2022cjo, Cabass:2023nyo, SimBIG:2023ywd, Stadler:2023hea, Ibanez:2023vxb, Nguyen:2024yth}, or using new summary statistics such as, for example, wavelets \cite{Valogiannis:2021chp, Eickenberg:2022qvy, Peron:2024xaw}.

The main results obtained in this paper can be summarized as follows: 
\begin{itemize}
\item we define the bootstrap approach more comprehensively than in \cite{DAmico:2021rdb}, by identifying explicitly all the requirements, from symmetry principles and other considerations, underlying the approach;
\item we derive the general structure of the displacement field independently on a specific form of the equations of motion explicitly up to sixth order in PT, and provide an algorithmic procedure to generate the contributions of arbitrary higher order;
\item we give examples, in specific models, of the equations of motion governing the time-dependence of the cosmology-dependent coefficients appearing in the general bootstrap expansion;
\item we consider modifications of GR of the Hordenski class of theories \cite{Horndeski:1974wa, Cusin:2017mzw, Cusin:2017wjg}, in which the Poisson equation is nonlinear, and verify that the general bootstrap expansion holds also in these cases;
\item we discuss the case in which dark matter is composed of different non-relativistic species, possibly with different initial densities and velocities.
\end{itemize}
In this paper we limit the discussion to dark matter. The case of biased tracers will be the subject of a future work.

The structure of this paper is as follows. In Sect.~\ref{sect:symmetries}, we identify the general requirements constraining the  structure of the displacement fields of $N_s$ non-relativistic matter species. In Sect.~\ref{sect:PT}, we  focus on building the generic terms that contribute to the displacement field in the case of single species, up to fourth order. After considering the extension to redshift space in Sect.~\ref{sect:RSD}, we  show how the general structure of the displacement fields are generated in GR in Sect.~\ref{LagRec}. In Sect.~\ref{sec:modgrav}, we show that the general expression for the displacement fields that we derived in the preceding sections also encompasses theories beyond general relativity, such as Horndeski theory. In Sect.~\ref{sec:Eul}, we discuss the relation between our approach and the eulerian bootstrap of Ref.~\cite{DAmico:2021rdb}. Lastly, in Sect.~\ref{sect:multispecies}, we extend this formulation to the case of matter  composed by different non-relativistic species. We will conclude with a summary and some future directions in Sect.~\ref{sec:conclusion}. 
Then, in Appendix \ref{app:psi_beyond_neq4},
we give the explicit results up to sixth order in LPT, and derive an algorithmic procedure to reach arbitrarily higher orders. In Appendix \ref{app:AppVort}
we derive explicitly the equation of motion for the vorticity component of the displacement field at fourth order. 
In Appendix \ref{app:MG} we give the explicit forms for the evolution equation for the bootstrap coefficient in the nDGP modified gravity model at third order.
Finally, in Appendix \ref{app:EPTmap} we give the mapping between the LPT bootstrap coefficients and the ones for the eulerian bootsrap approach of \cite{DAmico:2021rdb}.

\section{Constraints on the displacement fields}
\label{sect:symmetries} We consider $N_s$ non-relativistic matter species. The (eulerian) position $\bx$ at time $\tau$ of the particle of the $\alpha$ species ($\alpha=1,\cdots,N_s$) which was initially at the (lagrangian) position $\bq$ is given by
\beq
\bx=\bq+\bm{\psi}_\alpha(\bq,\tau)\,.
\label{eq:qtox}
\eeq
The displacement vector fields 
$\bm{\psi}_\alpha(\bq,\tau)$ will be the focus of our discussion. 
Assuming that the distribution of each species is uniform at initial conformal time, that is, 
\beq
\bm{\psi}_\alpha(\bq,\tau)\to 0\,,\qquad \mathrm{for}\;\;\tau\to 0\,,
\eeq
the overdensity for the $\alpha$-th species in eulerian space at time $\tau \ge 0$ is given by
\beq
\delta_\alpha(\bx,\tau)=\int d^3 q\,\delta_D\left(\bx-\bq-\bm{\psi}_\alpha(\bq,\tau)\right)-1\,.
\label{eq:psitodelta}
\eeq
Each displacement field admits a Helmholtz decomposition in terms of a scalar function, $\phi_\alpha(\bq,\tau)$, and a transverse (vorticity) vector field, $\bm{\omega}_\alpha(\bq,\tau)$,
\beq
\bm{\psi}_\alpha(\bq,\tau)=\bm{\psi}_{\alpha, s}(\bq,\tau)+\bm{\psi}_{\alpha, t}(\bq,\tau)=\bm{\nabla}\phi_\alpha(\bq,\tau)
+ \bm{\nabla}\times  \bm{\omega}_\alpha(\bq,\tau)\,,
\label{eq:Helmotz}
\eeq
with $\bm{\nabla}\cdot \bm{\omega}_\alpha=0$.
We will assume a perturbative expansion of the fields $\bm{\psi}_\alpha(\bq,\tau)$, $\phi_\alpha(\bq,\tau)$, and $\bm{\omega}_\alpha(\bq,\tau)$, 
\begin{align}
\bm{\psi}^N_\alpha(\bq,\tau)=&\sum_{n=1}^N  \bm{\psi}^{(n)}_\alpha(\bq,\tau)\,,\nonumber \\
\phi^N_\alpha(\bq,\tau)=&\sum_{n=1}^N  \phi^{(n)}_\alpha(\bq,\tau)\,,\nonumber\\
\bm{\omega}^N_\alpha(\bq,\tau)=&\sum_{n=1}^N  \bm{\omega}^{(n)}_\alpha(\bq,\tau)\,,
\label{eq:ptexp2}
\end{align}
where $N$ is the order of the truncation of the PT series.
The lowest, linear, order is defined by a scalar field (in general, one for each species), $\vpl_\alpha(\bq,\tau)$,
related to  the linear overdensity field, $\delta^{(1)}_\alpha(\bq,\tau)$,  by
\beq
\nabla^2 \vpl_\alpha(\bq,\tau)=-\delta^{(1)}_\alpha(\bq,\tau)\,.
\label{eq:vpldelta}
\eeq
At this order, the displacement fields and their longitudinal and transverse components are given by, 
\begin{align}
\bm{\psi}^{(1)}_\alpha(\bq,\tau)=&\,\bm{\nabla }\vpl_\alpha(\bq,\tau)\,,\nonumber\\
\phi^{(1)}_\alpha(\bq,\tau)=&\,\vpl_\alpha(\bq,\tau)\,,\nonumber\\
\bm{\omega}^{(1)}_\alpha(\bq,\tau)=&\,0\,,
\end{align}
In the case of a single species, and taking $\delta^{(1)}_\alpha(\bq,\tau)$ to be on the growing mode, inserting $\bm{\psi}^{(1)}_\alpha(\bq,\tau)$ in Eq.~\re{eq:psitodelta} gives the Zel'dovich approximation for the nonlinear density contrast.

At $n$-th order in LPT, the  displacement is built by combining $n$ linear scalar fields $\vpl_\alpha$. In the following, we discuss how various requirements constrain the form of these combinations.
\begin{itemize}
\item {\bf single stream regime:} in PT (both eulerian and lagrangian) we assume that the mapping between lagrangian and eulerian positions, given by  equation \re{eq:qtox},
is one-to-one \cite{Bernardeau:2001qr}\footnote{In eulerian PT, the single stream assumption corresponds  to neglecting velocity dispersion and all the higher order moments of the species distribution functions \cite{Pietroni:2011iz}.}.  As a consequence, the combination of linear fields giving the displacement field at a given LPT order is {\it local}, that is, all the fields are evaluated at the same lagrangian position $\bq$. We will discuss further this requirement in Sect.~\ref{sect:multispecies}, when we will deal with the multi-species case.
\item {\bf scale-independence:} since each $\vpl_\alpha$ has dimensions of [length]$^2$ (see Eq.~\re{eq:vpldelta}), and the displacement has dimension [length], $2n-1$ spatial derivatives should act on the $n$ linear fields. This counting is modified in presence of physical scales, such as the mass scale of new degrees of freedom in the gravity sector, or $k_{\rm NL}$, the wavevector scale above which perturbation theory assumptions, such as single streaming, break down. In this paper we will assume there are no new physical degrees of freedom, whereas $(k^2/k^2_{\rm NL})^m$ corrections are in general induced as UV counterterms  \cite{Baumann:2010tm, Pietroni:2011iz, Carrasco:2012cv}. 
\item{\bf rotational invariance:} the indices corresponding to $2n-2$ of the $2n-1$ spatial derivatives must be contracted by products of Kronecker delta functions and Levi-Civita tensors, leaving just one uncontracted index, corresponding to that of the vector displacement field.
\item {\bf Extended Galilean Invariance:} the dynamics is assumed \cite{Scoccimarro:1995if, Jain:1995kx, Peloso:2013zw, Kehagias:2013yd, Peloso:2013spa} to be invariant under a uniform time-dependent shift of all the particles' positions,
\beq
\bx_\alpha\to \bx_\alpha-\mathbf{d}(\tau)\,,\quad \forall \, \alpha\,,
\label{eq:gis}
\eeq
equal for all particle species, 
and a simultaneous redefinition of the gravitational force acting on them as
\beq
\mathbf{\nabla}_\bx \Phi_\alpha(\bx,\tau)\to \mathbf{\nabla}_\bx \Phi_\alpha(\bx,\tau) -\ddot{\mathbf{d}}(\tau) -{\cal H} \dot{\mathbf{d}}(\tau)\,,\quad \forall \, \alpha\,.
\label{eq:PhiEGI}
\eeq
In lagrangian space, the shift in Eq.~\re{eq:gis} corresponds to a common shift of the (fully nonlinear) displacement fields, 
\beq
\bm{\psi}_\alpha(\bq,\tau)\to \bm{\psi}_\alpha(\bq,\tau)+\mathbf{d}(\tau)\,,\quad \forall \, \alpha\,,
\label{eq:PsiEGI}
\eeq
at fixed lagrangian position $\bq$ \cite{Horn:2015dra}.
We require that the above invariance holds in any consistent approximation. Specifically, in LPT, a shift of the linear field,
\beq
\bm{\psi}^{(1)}_\alpha(\bq,\tau)\to \bm{\psi}^{(1)}_\alpha(\bq,\tau)+\mathbf{d}(\tau)\,,\quad \forall \, \alpha\,,
\label{eq:shiftlin}
\eeq
that is, 
\beq
\bm{\nabla }\vpl_\alpha(\bq,\tau)\to \bm{\nabla }\vpl_\alpha(\bq,\tau)+\mathbf{d}(\tau)\,,
\label{eq:shiftphi}
\eeq 
implies 
that the dynamics is such that the displacements  at  $N$-th order are shifted by the same amount,
\beq
\bm{\psi}^N_\alpha(\bq,\tau)\to \bm{\psi}^N_\alpha(\bq,\tau)+\mathbf{d}(\tau)\,,\quad \forall \, \alpha\,.
\eeq
Since the transformation is entirely encoded in the linear field, the nonlinear contributions to the displacement field 
must be invariant when the linear fields are shifted as in \re{eq:shiftphi},
\beq
\bm{\psi}^{(n)}_\alpha(\bq,\tau) \to \bm{\psi}^{ (n)}_\alpha(\bq,\tau)\,,\qquad n\ge 2\,.
\label{eq:EGIcon}
\eeq
This implies that linear fields $\vpl_{\alpha}$ can contribute to $\bm{\psi}^{(n)}_\alpha$  for $n\ge 2$ only in two ways, both invariant under the shift~\re{eq:shiftphi}:
\begin{itemize}
\item individually, as spatial derivatives at least of second order;
\item as first derivatives, in combinations such as
\beq
\bm{\nabla }\vpl_\alpha(\bq,\tau)-\bm{\nabla }\vpl_\beta(\bq,\tau)\,.
\label{eq:firstder}
\eeq
\end{itemize}
\item{\bf equivalence principle:} defining the average displacement for each species as
\beq
{\bm d}_\alpha(\tau)\equiv \frac{1}{V}\int d^3 q \,\bm{\psi}_\alpha(\bq,\tau)\,,
\eeq
where the integral is over the total lagrangian volume, $V$, 
then the configuration in which all of them and their time derivatives are equal,
\beq
{\bm d}_\alpha(\tau)={\bm d}(\tau)\,,\quad \dot{\bm d}_\alpha(\tau)=\dot{\bm d}(\tau)\,,\quad \forall\, \alpha\,,
\label{eq:fixpoint}
\eeq
is a {\it fixed point} of the dynamical evolution, that is, if it is realized at some time $\tau_0$, it will persist for all $\tau > \tau_0$. It implies that combinations containing first derivatives of the linear fields, such as those in Eq.~\re{eq:firstder}, are not generated by the purely gravitational dynamics. They can exist only as initial conditions or, at small scales, as the consequence of non-gravitational interactions, such as baryon-photon coupling, which induces a differential speed between baryon and cold dark matter \cite{Tseliakhovich:2010bj}.
Moreover, due to the expansion, the fixed point is attractive, meaning that initial large-scale velocity biases decay as the inverse of the scale factor, $\dot {\bm d}_\alpha-\dot{\bm d}_\beta\sim 1/a $.

\item {\bf mass conservation:} performing the integral in Eq.~\re{eq:psitodelta}, under the single streaming assumption gives
\beq
\delta_\alpha(\bx,\tau)=\frac{1}{J_\alpha(\bq_\alpha(\bx,\tau),\tau)}-1\,,
\label{eq:deltajac}
\eeq
where $J_\alpha(\bq_\alpha(\bx,\tau),\tau)$ is the Jacobian of the mapping from  lagrangian to eulerian coordinates,  which can be expressed as,
\begin{align}
J_\alpha(\bq,\tau)&=\det[\bm{1}+\bm{M}_\alpha(\bq,\tau)]\,, \nonumber\\
& =1+{ \rm Tr}[\bm{M}_\alpha(\bq,\tau)]+\frac{1}{2}\left({ \rm Tr}[\bm{M}_\alpha(\bq,\tau)]^2- { \rm Tr}[\bm{M}_\alpha(\bq,\tau)^2]\right)+\det[\bm{M}_\alpha(\bq,\tau)]\,,
\label{eq:det1pM}
\end{align}
evaluated on the solution of Eq.~\re{eq:qtox},
where $[M_{\alpha}(\bq,\tau)]^i_{\;j}\equiv \psi^i_{\alpha,j}(\bq,\tau)$ is the deformation tensor, and we introduced the notation,
\beq
(\cdots)_{,j}\equiv\frac{\partial}{\partial x^j}(\cdots)\,.
\eeq
Mass conservation for the species $\alpha$ is enforced if the integral of the density contrast vanishes at all times,
\begin{align}
\int d^3 x\,\delta_{\alpha}(\bx,\tau)& =\int d^3 q\, J_\alpha(\bq,\tau)\,\left(\frac{1}{J_\alpha(\bq,\tau)}-1\right)\,\nonumber\\
&=-\int d^3 q\, \left[\psi^i_{\alpha,i}+\frac{1}{2}\left(\psi^i_{\alpha,i}\psi^j_{\alpha,j}-\psi^i_{\alpha,j}\psi^j_{\alpha,i}\right)+\frac{1}{3!}\epsilon^{ikl}\epsilon^{jmn}\psi^i_{\alpha,j}\psi^k_{\alpha,m}\psi^l_{\alpha,n} \right]\,,\nonumber\\
&=-\int d^3 q\, \bigg[\nabla^2 \phi_\alpha +\frac{1}{2}\left(\phi_{\alpha,ii}\phi_{\alpha,jj}-\phi_{\alpha,ij}\phi_{\alpha,ji}\right)+ \frac{1}{3!}\epsilon^{ikl}\epsilon^{jmn}\phi_{\alpha,ij}\phi_{\alpha,km}\phi_{\alpha,ln} \nonumber\\
&\;\;\;\;\;\;\;\;\;\;\;\;\;\;\;\;\;\; -\frac{1}{2}\phi_{\alpha,ij}\left(\epsilon^{jmn}\omega^n_{\alpha,im}+ \epsilon^{imn}\omega^n_{\alpha,jm} \right) -\frac{1}{3!} \epsilon^{ikl}\epsilon^{jmn}\omega^l_{\alpha,jk}\,\omega^n_{\alpha,im}\nonumber\\
&\;\;\;\;\;\;\;\;\;\;\;\;\;\;\;\;\;\;+\frac{1}{3!}\epsilon^{ikl}\epsilon^{jmn}\bigg(
3\phi_{\alpha,ij}\phi_{\alpha,km} \epsilon^{lop} \omega^p_{\alpha,on} +3 \phi_{\alpha,ij}\epsilon^{kqr} \omega^r_{\alpha,qm} \epsilon^{lop} \omega^p_{\alpha,on}\nonumber\\
&\;\;\;\;\;\;\;\;\;\;\;\;\;\;\;\;\;\;+ 
\epsilon^{ist} \omega^t_{\alpha,js}\epsilon^{kqr} \omega^r_{\alpha,qm}\epsilon^{lop} \omega^p_{\alpha,on}\bigg)
\bigg]\nonumber\\
&=-\int d^3 q\, \psi^i_{\alpha,i}=-\int d^3 q \,\nabla^2 \phi_\alpha=0\,\,
\label{eq:masscon2}
\end{align}
where we have used Eq.~\re{eq:deltajac} and changed the integration variable. 
The  surviving terms at the last line are those that do not vanish upon integration by parts.
The condition Eq.~\re{eq:masscon2} is trivially verified at linear order but provides non-trivial constraints on the allowed form of higher order contributions to  the scalar component of the displacement field, $\phi_\alpha(\bq,\tau)$.
\item{\bf momentum conservation:} the momentum of the particle of species `$\alpha$' at the lagrangian coordinate $\bq$  is given by $\bp_\alpha=  m_\alpha a(\tau) \dot{\bm{\psi}}_\alpha(\bq,\tau)$, so, the total volume-averaged momentum is given by
\beq
\frac{1}{V}\int d^3 q\,\sum_\alpha \bar n_\alpha m_\alpha a(\tau) \dot{\bm{\psi}}_\alpha(\bq,\tau)=\Omega_m a(\tau) \sum_\alpha \omega_\alpha\, \dot{\bm{d}}_\alpha(\tau)\,,
\eeq 
where $\bar n_\alpha$ is the average particle number of the $\alpha$ species, and $\omega_\alpha\equiv \bar n_\alpha m_\alpha/\Omega_m=\Omega_\alpha/\Omega_m$ (that is, $\sum_\alpha \omega_\alpha=1$). EGI implies that the right hand side can always be set to zero by a shift of the linear fields as in Eq.~\re{eq:shiftlin}, with
\beq
\bm{d}(\tau)=\sum_\alpha \omega_\alpha \bm{d}_\alpha(\tau)\,,
\eeq
which does not affect the nonlinear contributions to the displacements. Therefore, 
momentum conservation implies that, order by order in LPT, we have 
\beq
\frac{1}{V}\int d^3 q\, \sum_\alpha \omega_\alpha \dot{\bm{\psi}}^{(n)}_\alpha(\bq,\tau)=0\,,\qquad n\ge 2\,,
\label{eq:momcon2}
\eeq
at all times. 
\end{itemize}
 In the next section we will discuss how all the conditions above constrain the nonlinear contributions to the displacement field. We will initially assume purely gravitational interactions
and initial conditions as in Eq.~\re{eq:fixpoint}, which effectively reduce the system to a single fluid.
Then, in Sect.~\ref{sect:multispecies}, we will highlight the modifications when large scale velocity bias is present and a multi-species treatment is needed.

\section{General structure of the nonlinear displacement}
\label{sect:PT}
Our purpose is to obtain the most general structure for the nonlinear contributions to the displacement field, that is, of the scalar and vector fields in Eq.~\re{eq:ptexp2} for $n\ge 2$. To reduce clutter, we will omit the space and time dependencies of the fields, when it will not cause ambiguities.

As discussed after Eq.~\re{eq:EGIcon}, in the single species case, EGI implies that the elementary building block of perturbation theory, namely, the linear scalar field $\vpl$, can appear only with at least two spatial derivatives,
\beq
\vpl_{,i j}\,.
\label{eq:vpl}
\eeq
Moreover, scale invariance implies that terms with a number of derivatives larger than two are not allowed.
Therefore, at $n$-th order we must find all possible independent combinations of the product of $n$ terms like Eq.~\re{eq:vpl} giving scalars and transverse vectors, once all the requirements discussed in the previous section have been imposed.
\subsection{Case $n=2$}
At second order, a  scalar field can be obtained as
\beq
\vp^{(2)}_{,i i}=\frac{1}{2}\left(\delta_{i_1 j_1}\delta_{i_2 j_2}+ c \,\delta_{i_1 j_2}\delta_{i_2 j_1} \right)\,\vpl_{,i_1 j_1}\vpl_{,i_2 j_2}\,,
\label{eq:vp2}
\eeq
where we sum over repeated indices and the constant $c$ is arbitrary, for now. 
Notice that, we defined the field $\vp^{(2)}$ so that it has the same dimensionality of $\vpl$ (namely, [lenght]$^2$), therefore it is its Laplacian who appears at the LHS. Mass conservation, Eq.~\re{eq:masscon2}, requires the integral of $\vp^{(2)}_{,i i}$ to vanish. 
Integrating twice by parts we get
\beq
\int d^3 q \,\vpl_{,i_1 j_1}\vpl_{,i_2 j_2}=-\int d^3 q\,\vpl_{, j_1}\vpl_{,i_1 i_2 j_2}=\int d^3 q \,\vpl_{,i_2 j_1}\vpl_{,i_1 j_2}\,,
\eeq
where, assuming periodic boundary conditions, the contributions from boundary terms vanish.
Therefore, to obtain a vanishing integral, the pre-factor in Eq.~\re{eq:vp2} must be antisymmetric under the exchange $i_1\leftrightarrow i_2$ or $j_1\leftrightarrow j_2$, which gives the only solution $c=-1$,
\begin{align}
\vp^{(2)}_{,i i}&\equiv\frac{1}{2}\left(\delta_{i_1 j_1}\delta_{i_2 j_2}- \,\delta_{i_1 j_2}\delta_{i_2 j_1} \right)\,\vpl_{,i_1 j_1}\vpl_{,i_2 j_2}=\frac{1}{2}\epsilon^{i_1 i_2 k}\epsilon^{j_1 j_2 k}\,\vpl_{,i_1 j_1}\vpl_{,i_2 j_2}\,,\nonumber\\
&=\frac{1}{2}\left(\vp^{(1)}_{,l l}\vp^{(1)}_{,j j}-\vp^{(1)}_{,l j}\vp^{(1)}_{,j l} \right)  \,,
\label{eq:vp22}
\end{align}
where $\epsilon^{ijk}$ is the Levi-Civita tensor.

At $n=2$ the only one viable scalar is then,
\beq
\phi^{(2)}(\bq,\tau)= c^{(2)}(\tau) \vp^{(2)}(\bq,\tau)\,, 
\label{eq:phi2}
\eeq
where the time dependent function $c^{(2)}(\tau)$ is determined, in a specific model, by the equations of motion (see Sects.~\ref{LagRec} and \ref{sec:modgrav}).
Notice that, since 
\beq
\vpl\equiv\phil \propto D(\tau) \,,
\eeq
where $ D(\tau)$ is the linear growth factor, the time dependence of the $n$-th order field $\vp^{(n)}$ is just
\beq
\vp^{(n)}\propto D^n(\tau)\,.
\eeq

At nonlinear orders we can also generate transverse vectors from lower order displacements, from combinations of the form,
\beq
\epsilon^{iln} \psi^{(n_1)\,j}_{a,l}\psi^{(n_2)\,j}_{b,n} \,,
\label{eq:nlvort}
\eeq
where the labels $a,\,b$ identify independent components of the displacements field of the same  order.
Notice that vectors built from three or more displacement fields, such as
\beq
\epsilon^{ilp} \psi^{(n_1)\,l}_{a,k}\psi^{(n_2)\,k}_{b,r}\psi^{(n_3)\,r}_{c,p} \,,
\eeq
do not vanish when integrated over the total volume, and are therefore forbidden by momentum conservation, Eq.~\re{eq:momcon2}.

In particular, to obtain a transverse vector from the product of two scalars, the only possible combination is
\beq
 \epsilon^{iln} \,\vp^{(n_1)}_{a,j l}\,\vp^{(n_2)}_{b,j n}\,.
\label{eq:transvect}
\eeq
Since at second order we can use only the linear field $\vpl$, we have $n_1=n_2=1$ and $a=b$, therefore the above term vanishes and no transverse vector is generated at this order.

The most general displacement field at second order is then
\beq
\bm{\psi}^{(2)}(\bq, \tau)= c^{(2)}(\tau) \bm{\nabla} \vp^{(2)}(\bq,\tau)\,. 
\eeq
\subsection{Case $n=3$}
\label{sect:n3}
The third order potential $\phi^{(3)}$ can be formed out of two types of contributions,
\beq
\vpl_{,i_1 j_1} \vpl_{,i_2 j_2}\vpl_{,i_3 j_3}\,,
\eeq
and
\beq
\vp^{(2)}_{,i_1 j_1} \vpl_{,i_2 j_2}\, \,.
\eeq
Mass conservation and rotational invariance require the indices to be contracted with a tensor antisymmetric under the exchange of two $i$'s or two $j$'s indices. We can then form two independent scalars,
\begin{align}
\vp_{1,ii}^{(3)}&\equiv \frac{1}{3!} \epsilon^{i_1 i_2 i_3}\,\epsilon^{j_1 j_2 j_3}\,\vpl_{,i_1 j_1} \vpl_{,i_2 j_2}\vpl_{,i_3 j_3}\,, \label{eq:vp31}\\
\vp_{2,ii}^{(3)}&\equiv \frac{1}{2!} \epsilon^{i_1 i_2 k}\,\epsilon^{j_1 j_2 k}\,\vp^{(2)}_{,i_1 j_1} \vpl_{,i_2 j_2}
\,. \label{eq:vp32}
\end{align}
Therefore the most generic third order scalar potential is
\beq
\phi^{(3)}(\bq,\tau)= c^{(3)}_1(\tau) \vp^{(3)}_1(\bq,\tau)+c^{(3)}_2(\tau) \vp^{(3)}_2(\bq,\tau)\,,
\eeq
where two new cosmology-dependent coefficients arise, $c^{(3)}_a(\tau)$ and $c^{(3)}_b(\tau)$.

At this order, we can also build a transverse vector. Indeed, we now can combine the two scalars $\vpl$ and $\vp^{(2)}$ as in Eq.~\re{eq:transvect} to get
\begin{equation}
    v_{,kk}^{(3)\, i} = \epsilon^{iln} \,\varphi^{(2)}_{,jl} \, \varphi^{(1)}_{,jn},  
\label{eq:vect3}
\end{equation}
and the vector component of the displacement field is given by
\beq
\omega^{(3)\, i}(\bq,\tau)=d^{(3)}(\tau) v^{(3) i}(\bq,\tau)\,,
\label{eq:vort3}
\eeq
swith  $d^{(3)}(\tau)$ a cosmology-dependent coefficient. The most general displacement field at third order is then
\beq
\bm{\psi}^{(3)}(\bq, \tau)= c^{(3)}_1(\tau) \bm{\nabla} \vp^{(3)}_1(\bq,\tau)+c^{(3)}_2(\tau) \bm{\nabla} \vp^{(3)}_2(\bq,\tau)+d^{(3)}(\tau) \bm{v}^{(3)}(\bq,\tau) \,. 
\eeq

The extension to higher orders is now clear. Since in three dimensions there are no antisymmetric tensors of order larger than three, the only possible terms are built using the tensors appearing in Eqs.~\re{eq:vp31} and \re{eq:vp32}. 

Then, at $n$-th order, we obtain the independent scalars by writing all possible combinations of derivatives of lower order displacements, of the form,
\beq
\vp^{(n)}_{a_n,ii} =\frac{1}{3!} \epsilon^{i_1 i_2 i_3}\,\epsilon^{j_1 j_2 j_3}\, \psi^{(m_1)\,i_1}_{a_{m_1},j_1}\psi^{(m_2)\,i_2}_{a_{m_2},j_2}\psi^{(m_3)\,i_1}_{a_{m_3},j_3}\,,
\label{eq:nl3}
\eeq
or 
\beq
\vp^{(n)}_{a_n,ii} =\frac{1}{2!} \epsilon^{i_1 i_2 k}\,\epsilon^{j_1 j_2 k}\, \psi^{(m_1)\,i_1}_{a_{m_1},j_1}\psi^{(m_2)\,i_2}_{a_{m_2},j_2}\,,
\label{eq:nl2}
\eeq
with $\sum m_i=n$.  Notice that the displacements at the RHS of the expressions above can be of the scalar or transverse type, as defined in \re{eq:Helmotz}.
When all of them are of the scalar type, the structure in Eqs.~\re{eq:nl3} and \re{eq:nl2} corresponds to the Galileon structures in three dimensions, see for instance, Refs.~\cite{Nicolis:2008in, Deffayet:2009wt, Deffayet:2009mn}. At lowest order, they give the terms in Eqs.~\re{eq:vp22}, \re{eq:vp31} and \re{eq:vp32}.

The independent transverse vectors are obtained from the combinations of the form \re{eq:nlvort},
\beq
v^{(n)\, i}_{,kk}=\epsilon^{iln} \psi^{(m_1)\,j}_{a_{m_1},l}\psi^{(m_2)\,j}_{a_{m_2},n}
\label{eq:nlv}
\eeq
and, in turn, they contribute to scalars at higher orders, through the structures in Eqs.~\re{eq:nl3}, ~\re{eq:nl2}, starting from $n=4$\footnote{As we will see in Sect.~\ref{sect:multispecies}, in the multi species case vectors can be generated already at $n=2$, by combining the linear displacements of the different species, so they start contributing to scalars at $n=3$.}.

The most general displacement field at $n$-th order is then obtained by a linear combination of gradients of the  independent scalars of the  form \re{eq:nl3}, \re{eq:nl2}, and of the independent vectors of type \re{eq:nlv}, each term being multiplied by a time and cosmology dependent coefficient.
\subsection{Case $n=4$}
At $n=4$ we can build  one scalar of the form Eq.~\re{eq:nl3}, by combining the second order scalar displacement \re{eq:vp22} and two linear ones,
\beq
\vp^{(4)}_{1,ii} \equiv \frac{1}{3!} \epsilon^{i_1 i_2 i_3}\,\epsilon^{j_1 j_2 j_3}\, \vp^{(2)}_{,i_1 j_1}\vp^{(1)}_{,i_2 j_2}\vp^{(1)}_{,i_3 j_3}\,,
\eeq
and three of the form Eq.~\re{eq:nl2},
\begin{align}
\vp^{(4)}_{2,ii} &\equiv \frac{1}{2!} \epsilon^{i_1 i_2 k}\,\epsilon^{j_1 j_2 k}\, \vp^{(3)}_{1,i_1 j_1}\vp^{(1)}_{,i_2 j_2}\,\nonumber\\
\vp^{(4)}_{3,ii} &\equiv \frac{1}{2!} \epsilon^{i_1 i_2 k}\,\epsilon^{j_1 j_2 k}\, \vp^{(3)}_{2,i_1 j_1}\vp^{(1)}_{,i_2 j_2}\,\nonumber\\
\vp^{(4)}_{4,ii} &\equiv \frac{1}{2!} \epsilon^{i_1 i_2 k}\,\epsilon^{j_1 j_2 k}\, \vp^{(2)}_{,i_1 j_1}\vp^{(2)}_{,i_2 j_2}\,,
\end{align}
where the first two contain the two independent scalar displacements at third order, Eqs.~\re{eq:vp31} and \re{eq:vp32}.

Moreover, we can combine the third order vector field, Eq.~\re{eq:vect3} with a linear scalar to get a fifth independent scalar field,
\beq
\vp^{(4)}_{5,ii} =\frac{1}{2}\vpl_{,lj}\left(\epsilon^{jmn} v^{(3)\,n}_{,lm}+ \epsilon^{lmn} v^{(3)\,n}_{,jm} \right)\,.
\eeq
A term of this form appears at the fourth line of  Eq.~\re{eq:masscon2}, and one can use integration by part and the transversality of the vector field to check that this term respects mass conservation.
In summary, the most generic scalar component of the displacement field at fourth order contains 5 independent contributions,
\beq
\phi^{(4)}(\bq,\tau)=\sum_{a=1}^5 c^{(4)}_a(\tau) \vp^{(4)}_a(\bq,\tau)\,.
\label{eq:vp4}
\eeq
As for the vectors, we now have two possible contributions from combining scalars,
\begin{align}
v_{1,kk}^{(4)\, i} &\equiv  \, \epsilon^{iln} \,\varphi^{(3)}_{1,jl}  \varphi^{(1)}_{,jn} \,,\nonumber\\
v_{2,kk}^{(4)\, i}&\equiv  \, \epsilon^{iln} \,\varphi^{(3)}_{2,jl}  \varphi^{(1)}_{,jn} \,,
\end{align}
and one further contribution involving the vector component of the third order displacement. From the general structure, Eq.~\re{eq:nlvort}, we get 
\beq
v_{3,kk}^{(4)\, i}(\bq, \tau) \equiv \epsilon^{iln}\,\epsilon^{jmp}\vpl_{,jl} v^{(3)\,p}_{,mn}\,,
\eeq
and therefore the most generic vector carries three cosmology dependent coefficients,
\beq
\omega^{(4)\, i}(\bq, \tau) =\sum_{a=1}^3 d^{(4)}_a(\tau) v_{a}^{(4)\, i}(\bq, \tau) \,.
\label{eq:vv4}
\eeq

This procedure can be extended to $n>4$. In Appendix \ref{app:psi_beyond_neq4}, we show that at fifth order there are 15 (11) independent scalar (vector) contributions to the displacement field, while at sixth order there are 52 (35) independent scalar (vector) contributions to the displacement field.
Moreover, we provide an algorithmic procedure to derive the available contributions at an arbitrarily higher order in LPT.

\section{Redshift space}
\label{sect:RSD}
The results of the previous section can be extended straightforwardly to redshift space. We will work in the plane parallel approximation in which the mapping from the lagrangian coordinate $\bq$ to the eulerian coordinate in redshift space, $\bs$, is given by 
\begin{equation}
    \mathbf{s} = \bx + \frac{(\bv\cdot \hat{\bz})\hat{\bz}}{\mathcal{H}} = \bq + \bm{\psi} + \frac{(\dot{\bm{\psi}}\cdot\hat{\bz})\hat{\bz}}{\mathcal{H}} \equiv \bq + \bm{\psi} +  \Delta_s \bm{\psi}\,,
\end{equation}
where the time derivative in the second equality is defined with respect to  conformal time. The redshift space  density contrast is given by the expression analogous to Eq.~\re{eq:psitodelta}, namely
\beq
\delta_{\rm rs}(\bs,\tau)=\int d^3 q\,\delta_D\left(\bs-\bq-\bm{\psi}(\bq,\tau)- \Delta_s \bm{\psi}(\bq,\tau)\right)-1\,.
\label{eq:psitodeltaRS}
\eeq
Using the expansions such as Eq.~\re{eq:vp4} and \re{eq:vv4}, the displacement to redshift space is then given at each perturbative order by
\begin{align}
 \Delta_s \bm{\psi}^{(n)\,i} = \left[\sum_{a=1}^{N_s(n)} c_a^{(n)} \left(n\, f +\frac{d \ln  c_a^{(n)}}{d \ln a}\right) \vp^{(n)}_{a,j}+  \sum_{a=1}^{N_v(n)} d_a^{(n)} \left(n\, f +\frac{d \ln  d_a^{(n)}}{d \ln a}\right) \epsilon^{jkl}v^{(n)\,l}_{a,k}\right] \hat z^j\,\hat z^i\,,
\label{eq:dpsirs}
\end{align}
where $f$  is the linear growth factor defined in Eq.~\re{eq:gammapar}, $N_s(n)$ ($N_v(n)$) is the number of independent scalar (vector) contributions at $n$-th order, and $c_1^{(1)}=1$.
Therefore, in redshift space, cosmology dependence enters not only through the coefficients $c_a^{(n)}$ and $d_a^{(n)}$, but also through their  derivatives with respect to the scale factor. This observation is very important, since when considering biased tracers, the only cosmology dependence not degenerate with the bias coefficients is precisely the one encoded in the velocity component of the displacement field, $\Delta_s \bm{\psi}^{(n),i}$, which is observationally accessible thanks to redshift space distortions via the Kaiser effect and its nonlinear generalizations \cite{Kaiser:1987qv, Scoccimarro:2004tg}.
Finally, notice that while in real space the vector components of the displacement field contribute to the density contrast starting from fourth order, in redshift space they contribute already at third order, thanks to the last term in Eq.~\re{eq:dpsirs}.

\section{Time dependence: GR}
\label{LagRec}
Now we show how the general expressions derived in the previous sections are generated in GR, where the role of DE is played by a cosmological constant, as in  $\Lambda$CDM, or by a dynamical fluid with time-dependent equation of state.

The equation of motion for the displacement field for a Newtonian pressureless self-gravitating fluid  in an expanding universe is
\beq
\frac{\partial^2 \bm{\psi}(\bq,\tau)}{\partial \tau^2}  + \mathcal{H}(\tau) \frac{\partial \bm{\psi}(\bq, \tau)}{\partial \tau} = -\left.\bm{\nabla}_x \Phi(\bx, \tau)\right|_{\bx=\bq + \bm{\psi}(\bq)}\,,
\label{eq:lageq2}
\eeq
where the gravitational potential in $\Lambda$CDM solves the Poisson equation
\beq
\nabla^2 \Phi(\bx,\tau)=\frac{3}{2}\mathcal{H}^2 \Omega_m(\tau) \delta(\bx,\tau)\,.
\eeq
One can verify directly that Eq.~\re{eq:lageq2} is invariant under EGI transfomations, Eqs.~\re{eq:PhiEGI} and \re{eq:PsiEGI}.
By taking the divergence and the curl of the equation of motion, we can derive the equations of motion for the scalar and the vector components of $\bm{\psi}$.

\subsection{Scalars}
Taking the divergence of Eq.~\re{eq:lageq2}, and using the Poisson equation and Eq.~\re{eq:deltajac}, we get
\beq
\nabla_\bx \frac{1}{a}\frac{\partial}{\partial \tau} \left(a \frac{\partial}{\partial \tau}\bm{\psi(\bq,\tau)}\right)=-\frac{3}{2}{\cal H}^2\Omega_m\left(\frac{1}{J(\bq,\tau)}-1\right)\,.
\label{eq:eomgrad}
\eeq
Then, using the relation,
\beq
\frac{\partial }{\partial x^i}=\frac{1}{2 J(\bq,\tau)} \epsilon_{ikl}\epsilon_{jmn}(\delta_{km}+\psi^k_{,m})(\delta_{ln}+\psi^l_{,n})\frac{\partial}{\partial q^j}\,,
\label{eq:gradx}
\eeq
we rewrite the equation as,

\beq
\frac{1}{2}\epsilon_{ikl}\epsilon_{jmn}(\delta_{km}+\psi^k_{,m})(\delta_{ln}+\psi^l_{,n})
 \frac{1}{a}\frac{\partial}{\partial \tau} \left(a \frac{\partial}{\partial \tau}\psi^i_{,j}\right)=\frac{3}{2}{\cal H}^2\Omega_m\left(J-1\right)\,,
 \label{eq:lapeq}
 \eeq
 which can be arranged as,
 \begin{align}
 &\frac{1}{a}\frac{\partial}{\partial \tau} \left(a \frac{\partial}{\partial \tau}\psi^i_{,i}\right)-\frac{3}{2}{\cal H}^2\Omega_m  \psi^i_{,i}  =-  \psi^i_{,i}
 \frac{1}{a}\frac{\partial}{\partial \tau} \left(a \frac{\partial}{\partial \tau}\psi^j_{,j}\right)+ \psi^i_{,j}
 \frac{1}{a}\frac{\partial}{\partial \tau} \left(a \frac{\partial}{\partial \tau}\psi^j_{,i}\right)\nonumber\\
 &-\frac{1}{2}\epsilon_{ikl}\epsilon_{jmn}\psi^k_{,m}\psi^l_{,n}\frac{1}{a}\frac{\partial}{\partial \tau} \left(a \frac{\partial}{\partial \tau}\psi^i_{,j}\right)+\frac{3}{2}{\cal H}^2\Omega_m \left[\frac{1}{2}\left( \psi^i_{,i}\psi^j_{,j}-\psi^i_{,j}\psi^j_{,i}\right)+\det\left(\psi^i_{,j}\right)\right]\,,
 \label{eq:nonlineareq}
\end{align}
where we have used Eq.~\re{eq:det1pM}.
The equation is still exact and, in its present form, it is particularly suited to a perturbative solution.

At linear order we have 
\beq
\frac{1}{a}\frac{\partial}{\partial \tau} \left(a \frac{\partial}{\partial \tau}\psi^{(1) \,i}_{,i}\right)-\frac{3}{2}{\cal H}^2\Omega_m  \psi^{(1)\, i}_{,i}=0\,.
\eeq
Since $\psi^{(1)\,i}_{,i}(\bq,\tau) =  \vpl_{,ii}(\bq,\tau) = D(\tau) \vpl_{,ii}(\bq,\tau=0)$, this is the  equation for the linear growth factor,
\beq
\frac{1}{a}\frac{\partial}{\partial \tau} \left(a \frac{\partial}{\partial \tau}D(\tau)\right)-\frac{3}{2}{\cal H}^2\Omega_m  D(\tau)=0\,.
\label{eq:Dlin}
\eeq
At second order, using  $\psi^{(1)\,i}_{,j}(\bq,\tau)=  \vpl_{,ij}(\bq,\tau)$, we get
\begin{align}
\frac{1}{a}\frac{\partial}{\partial \tau} \left(a \frac{\partial}{\partial \tau}\psi^{(2)\,i}_{,i}\right)-\frac{3}{2}{\cal H}^2\Omega_m\psi^{(2)\,i}_{,i}=&-\frac{3}{2}{\cal H}^2\Omega_m \left(\vpl_{,ii}\vpl_{,jj}-\vpl_{,ij}\vpl_{,ji} \right)\nonumber\\
&+\frac{3}{2}{\cal H}^2\Omega_m\left(\frac{1}{2}\left(\vpl_{,ii}\vpl_{,jj}-\vpl_{,ij}\vpl_{,ji} \right) \right)\,.
\end{align}
Using Eq.~\re{eq:phi2}, we have $\psi^{(2)\,i}_{,j}(\bq,\tau)= c^{(2)}(\tau) \vp^{(2)}_{,ij}(\bq,\tau)$, which, inserted in the previous equations gives the equation for $c^{(2)}(\tau)$,
\beq
\frac{1}{a D^2}\frac{\partial}{\partial \tau} \left(a \frac{\partial}{\partial \tau}(c^{(2)}(\tau)D^2)\right)-\frac{3}{2}{\cal H}^2\Omega_m(c^{(2)}(\tau)-1)=0\,.
\label{eq:c2eq}
\eeq
At third order we get,
\beq
\frac{1}{a}\frac{\partial}{\partial \tau} \left(a \frac{\partial}{\partial \tau}\psi^{(3)\,i}_{,i}\right)-\frac{3}{2} \mathcal{H}^2\Omega_m 
\psi^{(3)\,i}_{,i}= \frac{3}{2} \mathcal{H}^2\Omega_m \left[
-2 \det (\psi^{(1)\,i}_{,j}) +2(1-c^{(2)}) \frac{  \psi_{,i}^{(2)\,i} \psi_{,j}^{(1)\,j} - \psi_{,j}^{(2)\,i} \psi_{,i}^{(1)\,j} }{2}\right]\,, 
\eeq
from which we read the equations of motion for $c^{(3)}_{1,2}$, namely,
\begin{align}
\frac{1}{a D^3}\frac{\partial}{\partial \tau} \left(a \frac{\partial}{\partial \tau}(c^{(3)}_1(\tau)D^3)\right)&-\frac{3}{2}{\cal H}^2\Omega_m\left( c^{(3)}_1(\tau) -2 \right)=0\,,\nonumber\\
\frac{1}{a D^3}\frac{\partial}{\partial \tau} \left(a \frac{\partial}{\partial \tau}(c^{(3)}_2(\tau)D^3)\right)&-\frac{3}{2}{\cal H}^2\Omega_m\left( c^{(3)}_2(\tau) +2 (1-c^{(2)}(\tau) )\right)=0\,.
\label{eq:c3eq}
\end{align}
In the EdS limit ($\Omega_m=1$, $a\sim \tau^2$, ${\cal H}=2/\tau$) we get the constant solutions $\{c^{(2)},c^{(3)}_1,c^{(3)}_2\}=\{-3/7,\,-1/3,\,10/21\}$. Notice that evolving dark energy models modify Eqs.~\eqref{eq:c2eq} and \eqref{eq:c3eq} only through the background evolution, as in the $w_0$-$w_a$ parametrization \cite{CP2003, Linder_wa},
\begin{equation}
    {\cal H}(a) = a\,H_0 \sqrt{\Omega_{\rm m}(a/a_0)^{-3} + (1-\Omega_{\rm m}) (a/a_0)^{-3 (1+w_0+w_a)} e^{3 w_a (a/a_0-1)}} \;.
\end{equation}

\subsection{Vorticity}
\label{sect:vorto}
Following, for example \cite{Matsubara:2015ipa}, we take the curl of Eq.~\re{eq:lageq2}
and use again Eq.~\re{eq:gradx}, to obtain the equation for the transverse vector field,
\beq
\frac{1}{a}\frac{\partial}{\partial \tau}\left( a \frac{\partial}{\partial \tau}\omega^i_{,mm}\right)  = -\epsilon^{ijk}\,\psi^l_{, j} \frac{1}{a}\frac{\partial}{\partial \tau}\left( a \frac{\partial}{\partial \tau}\psi^l_{,k}\right)\,.
\label{eq:vorttot}
\eeq
Notice that this equation is independent on the gravitational force, the only assumption being that it is the gradient of a scalar potential. Therefore, its form is independent on the cosmological model and, moreover, the vorticity coefficients are  not independent on  the scalar ones, as we will see below.

As seen in Sect.~\ref{sect:PT}, vorticity starts to be generated at third order,
\beq
\frac{1}{a}\frac{\partial}{\partial \tau} \left(a \frac{\partial}{\partial \tau} \omega^{(3)\,i}_{,mm}  \right) = - \epsilon^{ijk} \left[ \varphi^{(2)}_{,lj} \, \frac{1}{a}\frac{\partial}{\partial \tau} \left(a \frac{\partial}{\partial \tau} \varphi^{(1)}_{,lk} \right) -  \frac{1}{a}\frac{\partial}{\partial \tau} \left(a \frac{\partial}{\partial \tau}\, \varphi^{(2)}_{,lj} \right) \varphi^{(1)}_{,lk}
\right]\,.
\eeq
Using Eq.~\re{eq:vort3}, we get 
the following equation of motion for the coefficient $d^{(3)}(\tau)$,
\begin{align}
\frac{1}{a D^3 }\frac{\partial}{\partial \tau} \left(a \frac{\partial}{\partial \tau} \left(d^{(3)}(\tau) D^3 \right)  \right) & = c^{(2)}(\tau) \frac{1}{a}\frac{\partial}{\partial \tau} \left(a \frac{\partial}{\partial \tau}D(\tau)\right)  - \frac{1}{a D^2}\frac{\partial}{\partial \tau} \left(a \frac{\partial}{\partial \tau}(c^{(2)}(\tau)D^2)\right)\,,
\nonumber\\
&= \frac{3}{2} \mathcal{H}^2 \Omega_m,
\end{align}
where, in the last line, we used the equations of motion for the scalar coefficients in $\Lambda$CDM, Eqs.~\re{eq:Dlin} and \re{eq:c2eq}. Notice, however, that the first equation holds in any model respecting the conditions listed in Sect.~\ref{sect:symmetries},
and shows that, as anticipated, the vorticity coefficient is not an independent parameter, being  entirely determined by the growth factor $D(\tau)$ and by the second order scalar coefficient, $c^{(2)}(\tau)$.
In the EdS limit, we have $d^{(3)}(\tau) = 1/7$.

In Appendix \ref{app:AppVort} we give the equations of motion for the vorticity coefficients at fourth order, which are, again, entirely determined by the  scalar and vector coefficients at lower orders.

\section{Time dependence: modified gravity}
\label{sec:modgrav}

In this section, we consider the equations of motion in theories beyond general relativity. First, let us show that no new contributions to the nonlinear displacement are generated with respect to the ones derived in Sect.~\ref{sect:PT} and present in $\Lambda$CDM. In particular, we consider models with an extra scalar field coupled to gravity, described by the Horndeski lagrangian \cite{Horndeski:1974wa}, the most general scalar-tensor theory yielding  second order field equations. Besides the new scalar field, $\chi$, the equations of motion involve 
 the scalar perturbations $\Phi$ and $\Psi$, defined from 
\begin{equation}
ds^2 = - (1+2 \Phi)dt^2 + a^2(t) (1-2 \Psi) \delta_{ij} dx^i dx^j.
\end{equation}

As shown in \cite{Cusin:2017mzw, Cusin:2017wjg}, in the quasi-static limit, where the dominating operators in the action are the ones with the highest number of spatial derivatives in the fields, the Poisson equations are modified into
\begin{align}
\label{eq:PoissonHorOri}
A_{da} \nabla_x^2 \Phi_a(\bx, \tau) = 
\frac{3}{2} \mathcal{H}^2 \Omega_m \delta_{d1} \delta(\bx, \tau) &+ \frac{B_{dab}}{4 \mathcal{H}^2} \epsilon_{ikm} \epsilon_{jlm} \frac{\partial^2 \Phi_a (\bx, \tau)}{\partial x_i \partial x_j} \frac{\partial^2 \Phi_b (\bx, \tau)}{\partial x_k \partial x_l} 
\nonumber\\ &+
\frac{C_{dabc}}{12 \mathcal{H}^2} \epsilon_{ikm} \epsilon_{jln} \frac{\partial^2 \Phi_a (\bx, \tau)}{\partial x_i \partial x_j} \frac{\partial^2 \Phi_b (\bx, \tau)}{\partial x_k \partial x_l} \frac{\partial^2 \Phi_c (\bx, \tau)}{\partial x_m \partial x_n},
\end{align}
where we have used a compact notation, with $\Phi_a \equiv ( \Phi, \Psi, \chi)$ for $a = 1, 2, 3$, and some coupling coefficients $A_{da}, B_{dab}, C_{dabc}$. These equations can be diagonalized as follows
\begin{align}
\label{eq:PoissonHor}
\nabla_x^2 \Phi(\bx, \tau) = 
\frac{3}{2} \alpha \mathcal{H}^2 \Omega_m \delta(\bx, \tau) &+ \frac{B'_{ab}}{4 \mathcal{H}^2} \epsilon_{ikm} \epsilon_{jlm} \frac{\partial^2 \Phi_a (\bx, \tau)}{\partial x_i \partial x_j} \frac{\partial^2 \Phi_b (\bx, \tau)}{\partial x_k \partial x_l} 
\nonumber\\ &+
\frac{C'_{abc}}{12 \mathcal{H}^2} \epsilon_{ikm} \epsilon_{jln} \frac{\partial^2 \Phi_a (\bx, \tau)}{\partial x_i \partial x_j} \frac{\partial^2 \Phi_b (\bx, \tau)}{\partial x_k \partial x_l} \frac{\partial^2 \Phi_c (\bx, \tau)}{\partial x_m \partial x_n},
\end{align}
for some coefficients $\alpha, B'_{ab}, C'_{abc}$. Let us denote the last two terms of Eq.~\eqref{eq:PoissonHor} as $\Delta \nabla_x^2 \Phi(\bx, \tau)$.

From Eq.~\eqref{eq:lageq2}, \re{eq:eomgrad}, and \re{eq:gradx}, these extra terms will contribute to the lagrangian equation of motion as $J(\bq) \Delta \nabla_x^2 \Phi(\bx, \tau)$.  Integrating in $d^3q$, we get,
\begin{align}
\int d^3 q ~J(\bq) \Delta \nabla_x^2 \Phi(\bx, \tau) = 
\int d^3 x & \bigg[ \frac{B'_{ab}}{4 \mathcal{H}^2} \epsilon_{ikm} \epsilon_{jlm} \frac{\partial^2 \Phi_a (\bx, \tau)}{\partial x_i \partial x_j} \frac{\partial^2 \Phi_b (\bx, \tau)}{\partial x_k \partial x_l} \nonumber \\
&+
\frac{C'_{abc}}{12 \mathcal{H}^2} \epsilon_{ikm} \epsilon_{jln} \frac{\partial^2 \Phi_a (\bx, \tau)}{\partial x_i \partial x_j} \frac{\partial^2 \Phi_b (\bx, \tau)}{\partial x_k \partial x_l} \frac{\partial^2 \Phi_c (\bx, \tau)}{\partial x_m \partial x_n} \bigg]
= 0,
\end{align}
where the last equality follows from integration by parts. Therefore, the extra terms allowed by the Horndeski theories, $J(\bq) \Delta \nabla_x^2 \Phi(\bx, \tau)$, vanish when integrated over the lagrangian space. This is the same requirement that the we enforced in Eq.~\eqref{eq:masscon2}, when building the general scalar contributions to the displacement fields. Hence, we conclude that in the Horndeski theories, no new type of terms are being generated at any order in perturbation theory. 

\begin{figure}
    \centering
    \includegraphics[width=\textwidth]{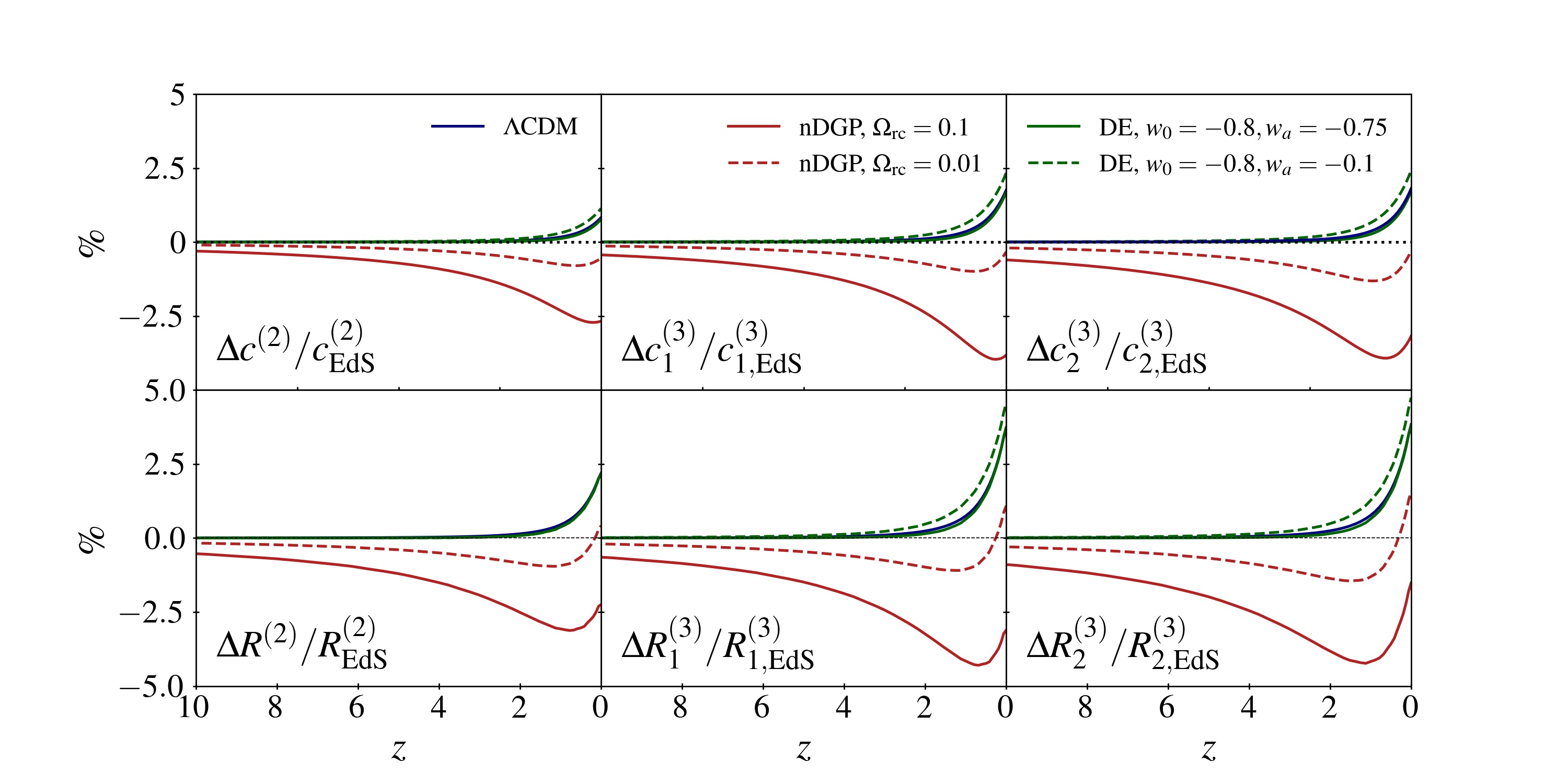}
    \caption{\textbf{Upper plots}: real space function for matter, $c^{(2)}, c_1^{(3)}$ and $c_2^{(3)}$ as a function of redshift. We show their time dependence for different cosmological model, $\Lambda$CDM with $\Omega_m$ fixed at the Planck best value, the modified gravity theory nDGP, and dark energy model with evolving equation of state parametrized with $w_0$-$w_a$. Since evolving dark energy modifies only the background evolution, the deviations of these model from $\Lambda$CDM is much smaller compared to theories that modifies the time dependence of the perturbations, such as modified gravity theories. \textbf{Lower plots}: we show the analogous for the redshift space coefficients,s $R^{(2)} \equiv 2c^{(2)} + \partial_\chi c^{(2)}$, $R_1^{(3)}~\equiv~2 c_1^{(3)} +\partial_\chi c_1^{(3)}$ and $R_2^{(3)} \equiv 2 c_2^{(3)} + \partial_\chi c_2^{(3)}$, see main text.}
    \label{fig:realc}
\end{figure}

Now, let us discuss a particular model beyond general relativity. In the renowned modified gravity model nDGP~\cite{Dvali:2000hr} model, Eq.~\eqref{eq:c2eq} becomes, explicitly,
\begin{equation}
    \partial_\chi^2 c^{(2)} + \left(3 + \frac{3}{2}\frac{\Omega_m}{f^2}\mu\right) \partial_\chi c^{(2)} + \left(2 + \frac{3}{2}\frac{\Omega_m}{f^2}\mu\right)c^{(2)} + \frac{3}{2}\frac{\Omega_m}{f^2}\mu - 2 \frac{\mu_2}{f^2}\left(\frac{3}{2}\Omega_m\right)^2 = 0\,,
\end{equation}
where we have introduce the more suitable time variable $d\chi \equiv d\log{D}$, and the $\mu$ and $\mu_2$ are time dependent functions parametrizing, respectively, deviations from GR at linear and second order, reported in App.~\ref{app:MG}.  Following~\cite{Piga:2022mge}, we assume a $\Lambda$CDM background. The nDGP version of the third order coefficients, Eq.~\eqref{eq:c3eq}, can be found in App.~\ref{app:MG}. We plot the time evolution of the bootstrap coefficients in Fig.~\ref{fig:realc} in three case, for $\Lambda$CDM, $w_0w_a$CDM, and nDGP.
These effects are different in size and sign for the different models, and can therefore break the degeneracy of analyses based on measurements of linear growth factor $f$ only.

\section{Relation with eulerian bootstrap}
\label{sec:Eul}

In this section we will discuss the relation between the present approach and the eulerian bootstrap
approach, derived in~\cite{DAmico:2021rdb} and used in the analyses~\cite{Amendola:2023awr, Amendola:2024gkz}.
In eulerian perturbation theory (EPT) the density contrast is represented as an expansion
\beq
\label{expansion}
\delta^N (\bx,\tau) = \sum_{n=1}^N \,\delta^{(n)} (\bx,\tau), 
\eeq
where the $n$-th order term, in Fourier space, is given by a convolution of the linear fields,
\beq
\delta^{(n)}_\bk (\tau)=\frac{1}{n!}\int \frac{d^3\bp_1}{(2\pi)^3}\cdots \frac{d^3\bp_n}{(2\pi)^3}\,(2\pi)^3 \delta(\bk-\bp_{1\cdots n})\, F^{(n)}(\bp_1,\,\cdots, \bp_n;\tau)\,\delta^{(1)}_{\bp_1} (\tau)\cdots \delta^{(1)}_{\bp_n} (\tau)\,,
\label{eq:EPTexp}
\eeq
with $\bp_{1\cdots n}\equiv \sum_{i=1}^n\bp_i$. An analogous expression holds for the velocity divergence at $n$-th order, $\theta^{(n)}_{\bk}(\tau)$,  with different kernels, indicated as $G^{(n)}$. The EPT kernels are determined, in each model, by an iterative solution of the system of continuity, Euler, and Poisson equations, see for instance \cite{Bernardeau:2001qr}.
In the EPT bootstrap approach \cite{DAmico:2021rdb}, on the other hand, one does not consider equations of motion, but expresses the kernels as the most general sum of rotational invariant combinations of the momenta $\bp_1,\,\cdots,\bp_n$, and then reduces the number of independent coefficients by imposing a set of constraints equivalent to the ones listed in Sect.~\ref{sect:symmetries}.
While the philosopy is the same as for the LPT bootstrap approach introduced in this paper, the practical implementation turns out to be less agile. The main reason for that is that the EPT bootstrap is based on the `destruction' of all the operator combinations not compatible with the constraints. The number of initial independent terms, and the number of constraints to impose increase rapidly with the perturbative order, so the procedure becomes more and more time-consuming as higher orders are considered. On the other hand, LPT bootstrap is based on the `construction' of the allowed operators starting from second derivatives of the linear field compatible with a  set of rules which does not change with the PT order, and is therefore naturally implemented in an algorithmic procedure.

The different agility of the two approaches emerges more clearly if one considers EGI constraints. In LPT the basic quantity is the displacement field, defined in lagrangian space. The effect of a uniform shift of the eulerian coordinates on the displacement field is just the additive transformation \re{eq:PsiEGI} which, in turn, implies that all the nonlinear contributions are built up by second derivatives of the linear field. 
On the other hand, in EPT the fundamental quantity is the density  field, which is defined in eulerian space. It  transforms nonlinearly under a uniform shift of eulerian coordinates,
\beq
\delta(\bx,\tau)\to \delta(\bx-\bm{d}(\tau),\tau)\,,
\eeq
which in Fourier space gives
\beq
\delta_\bk(\tau)\to e^{i \bk\cdot \bm{d}(\tau)}\delta_\bk(\tau)\,.
\eeq
As a consequence of the nonlinearity of the transformation, EGI constraints in EPT connect kernels at different PT orders \cite{DAmico:2021rdb}. The simplest, leading order (LO), forms of these relations constrain the limit of the EPT kernels when some of the external momenta go to zero individually,
\beq
\lim_{\bp_1,\cdots,\bp_m\to 0}F^{(n)}(\bp_1,\cdots,\bp_n,\tau)= \frac{\bp_{1}\cdot\bp_{1\cdots n}}{p_1^2}\cdots \frac{\bp_{m}\cdot\bp_{1\cdots n}}{p_m^2}F^{(n-m)}(\bp_{m+1},\cdots,\bp_n,\tau)\,.
\label{eq:LOEGI}
\eeq
Then, starting at third order, there are `next to leading order' (NLO) constraints, on the limit of the kernels when the sum of two momenta go to zero. From fourth order one has to impose also NNLO constraints, when the sum of three momenta goes to zero, and so on.  
All these constraints can be easily rederived starting from the expression of the density field in terms of the displacement, Eq.~\re{eq:psitodelta}, which, in Fourier space reads (we discuss for simplicity here the real space case),
\begin{equation}
    \delta(\bk,\tau) = \int \, d^3\,\bq e^{i \bk\cdot \bq}e^{i \bk\cdot \bm{\psi}(\bq,\tau)} - (2\pi)^3\delta_D(\bk)\,,
    \label{eq:ptexpf}
\end{equation}
by first expressing the displacement field in LPT, Eq.~\re{eq:ptexp2}, and then expanding the exponential up to a given order in the linear fields,  to recover the EPT expansion \re{eq:EPTexp}. For instance, the constraint in \re{eq:LOEGI} is obtained from the contribution to the exponential in \re{eq:ptexpf} of the form
\beq
\frac{\left(i \bk \cdot \bm{\psi}^{(1)}(\bq,\tau)\right)^m}{m!}\,\left[ e^{i \bk\cdot \left(\sum_{n=2}\bm \psi^{(n)}(\bq,\tau)\right)}\right]_{n-m}\,,
\eeq
where $[\cdots]_{n}$ indicates that one has to expand the quantity inside parentheses and keep the contributions of $n$-th order. NLO constraints correspond to expansions of \re{eq:ptexpf} in powers of $\bm{\psi}^{(2)}$, and so on. In other words, all the EGI constraints of EPT are just a consequence of the LPT constraints on the displacement field and of the relation between $\bm{\psi}$ and $\delta$, Eq.~\re{eq:ptexpf}.

In appendix \ref{app:EPTmap} we derive the explicit mapping between the EPT bootstrap parameters and the LPT ones, up to third order.

\section{Multispecies}
\label{sect:multispecies}
We now consider the case in which matter is composed by different non-relativistic species, each described by a different displacement field. A prominent situation for  which  this is relevant is the description of the baryon-cold dark matter system, where relative displacements and velocities are imprinted at recombination and after reionization by non gravitational forces such as Compton drag and baryonic pressure \cite{Tseliakhovich:2010bj, Chen:2019cfu, Rampf:2020ety, Hahn:2020lvr, Lewandowski:2014rca, Braganca:2020nhv}. Another example is the different clustering properties of massive neutrinos and  cold dark matter \cite{Aviles:2020cax}.

We consider, for simplicity, two cold dark matter species, `$\alpha$' and `$\beta$' with relative  abundances $\omega_\alpha$ and $\omega_\beta$ ($\omega_\alpha+\omega_\beta=1$). At a given time $\tau$, the same eulerian point $\bx$ is occupied by particles of the two species which were initially at the two, generally different, positions $\bq_\alpha(\bx,\tau)$ and  $\bq_\beta(\bx,\tau)$, given by the solutions of the equations
\beq
\bx=\bq_\alpha+\bm{\psi}_\alpha(\bq_\alpha,\tau)=\bq_\beta+\bm{\psi}_\beta(\bq_\beta,\tau)\,.
\label{eq:qbtoqa2}
\eeq
Since our approach is perturbative, we must assume that both mappings, $\bq_\alpha \leftrightarrow \bx$ and $\bq_\beta \leftrightarrow \bx$, are invertible, see the discussion in Sect.~\ref{sect:symmetries}. This implies that also the mapping $\bq_\alpha \leftrightarrow \bq_\beta$ is invertible. As anticipated after Eq.~\re{eq:EGIcon}, in the multispecies case we can build EGI invariant terms not only from spatial derivatives of the displacements (that is, second derivatives of the scalar and vector fields) but also from differences between displacement fields,
\beq
\bm{\psi}_\alpha^{(n)\,i}- \bm{\psi}_\beta^{(n)\,i}\,,
\eeq
as they are invariant under a common uniform shift of the two displacements. 
Moreover, mass conservation should hold independently for the $\alpha$ and the $\beta$ species,
\begin{align}
0=\int d^3 x\,\delta_\alpha(\bx) &= \int d^3 q_\alpha \left(1-J_\alpha(\bq_\alpha)\right)\,,\nonumber\\
0=\int d^3 x\,\delta_\beta(\bx) &=\int d^3 q_\beta \left(1-J_\beta(\bq_\beta)\right)=\int d^3q_\alpha \left(J_{\beta\alpha}(\bq_\alpha)-J_\alpha(\bq_\alpha) \right)\,,
\end{align}
where $J_{\beta\alpha}(\bq_\alpha)$ is the jacobian of the transfomation $\bq_\alpha \to \bq_\beta$. While the terms in $J_\alpha(\bq_\alpha)$  give the same constraints on the nonlinear contributions to the displacement $\bm{\psi}_\alpha^{i}$ than Eq.~\re{eq:masscon2}, analyzed in Sect.~\ref{sect:PT} (and analogously for $\bm{\psi}_\beta^{i}$), the new term in $J_{\beta\alpha}(\bq_\alpha)$ generates genuinely multi-species terms. Indeed, 
writing the solution of Eq.~\re{eq:qbtoqa2} as
\beq
\bq_\beta(\bq_\alpha,\tau)=\bq_\alpha+\Delta \bq_{\beta\alpha}(\bq_\alpha,\tau)\,,
\eeq
we have
\beq
J_{\beta\alpha}(\bq_\alpha)=\det\left(\delta^{ij}+\Delta q^i_{\beta\alpha,j}\right)= 1+{\rm Tr}\left[\Delta Q_{\beta\alpha}\right]+\frac{1}{2}\left({\rm Tr}\left[\Delta Q_{\beta\alpha}\right]^2-{\rm Tr}\left[\Delta Q_{\beta\alpha}^2\right]\right)+\det\left[\Delta Q_{\beta\alpha}\right],
\label{eq:detbetaalpha}
\eeq
where 
\beq
\left[\Delta Q_{\beta\alpha}\right]_{ij}\equiv \Delta q^i_{\beta\alpha,j}\,.
\eeq
Solving Eq.~\re{eq:qbtoqa2} perturbatively up to third order, we get
\begin{align}
\left[\Delta Q_{\beta\alpha}\right]_{ij} &=\left(\psi^{(1)\,i}_{\alpha,j}-\psi^{(1)\,i}_{\beta,j}\right)+\left(\psi^{(2)\,i}_{\alpha,j}-\psi^{(2)\,i}_{\beta,j}\right)-\left[\psi^{(1)\,i}_{\beta,l}\left( \psi^{(1)\,l}_{\alpha}-\psi^{(1)\,l}_{\beta} \right)\right]_{,j}
\nonumber\\&
+\left(\psi^{(3)\,i}_{\alpha,j}-\psi^{(3)\,i}_{\beta,j}\right) - \left[\psi^{(1)\,i}_{\beta,l}\left( \psi^{(2)\,l}_{\alpha}-\psi^{(2)\,l}_{\beta} - \psi^{(1)\,l}_{\beta,k}\left( \psi^{(1)\,k}_{\alpha}-\psi^{(1)\,k}_{\beta} \right)\right)\right]_{,j}
\nonumber\\&
- \left[\psi^{(2)\,i}_{\beta,l}\left( \psi^{(1)\,l}_{\alpha}-\psi^{(1)\,l}_{\beta}\right) \right]_{,j} 
- \frac{1}{2}\left[\psi^{(1)\,i}_{\beta,lk}\left( \psi^{(1)\,l}_{\alpha}-\psi^{(1)\,l}_{\beta}\right) \left( \psi^{(1)\,k}_{\alpha}-\psi^{(1)\,k}_{\beta}\right) \right]_{,j}
\nonumber\\& 
+\cdots,
\label{eq:DQ}
\end{align}
where all terms are evaluated at the same lagrangian position.
Taking the trace, as in the second term at the RHS of Eq.~\re{eq:detbetaalpha}, we get, already at second order,  a  scalar contribution involving the two displacements,
\beq
\vp^{(2),{\rm rel}}_{\beta \alpha,kk}\equiv \left[\psi^{(1)\,i}_{\beta,l}\left( \psi^{(1)\,l}_{\alpha}-\psi^{(1)\,l}_{\beta} \right)\right]_{,i}=\left[\vpl_{\beta,il}\left( \vpl_{\alpha,l}-\vpl_{\beta,l} \right)\right]_{,i}\,.
\label{eq:diff2}
\eeq
Unlike the terms discussed in  Sect.~\ref{sect:PT}, it contains first derivatives of scalar fields. Moreover, being a total derivative, it vanishes upon integration in $\bq_\alpha$. The same happens for all the third order terms of the new type in Eq.~\re{eq:DQ}. 

To see explicitly how the new terms are generated, we consider a $\Lambda$CDM model with two species.
The Poisson equation  reads
\beq
\nabla_\bx^2 \Phi = \frac{3}{2}{\cal H}^2\Omega_m \left(\omega_\alpha \delta_\alpha +\omega_\beta \delta_\beta \right)\,,
\eeq
and the  equations of motion for the displacement fields are,
\begin{align}
\frac{1}{a}\frac{\partial}{\partial \tau}\left(a \frac{\partial}{\partial \tau} \bm{\psi}_\alpha(\bq_\alpha)\right) &=-\left.\bm{\nabla}_\bx \Phi(\bx)\right|_{\bx=\bq_\alpha+\bm{\psi}_\alpha(\bq_\alpha)}\,,\label{eq:psia22}\\
\frac{1}{a}\frac{\partial}{\partial \tau}\left(a \frac{\partial}{\partial \tau} \bm{\psi}_\beta(\bq_\beta)\right) &=-\left.\bm{\nabla}_\bx \Phi(\bx)\right|_{\bx=\bq_\beta+\bm{\psi}_\beta(\bq_\beta)}\,.
\end{align}
Both equations are invariant under EGI transfomations, Eq.~\re{eq:PhiEGI}, ~\re{eq:PsiEGI},
and each of them conserves mass independently. Indeed, taking the gradient as in Eq.~\re{eq:eomgrad}, we get, for the first equation, 
\begin{align}
0=\int d^3 x \, \nabla_\bx^2 \Phi(\bx) =&
\int d^3 q J_\alpha(\bq) \frac{\partial}{\partial x^i}  \frac{1}{a}\frac{\partial}{\partial \tau}\left(a \frac{\partial}{\partial \tau} \psi^i_\alpha(\bq)\right)\,,\nonumber\\
=& \frac{1}{a}\frac{\partial}{\partial \tau}\left(a \frac{\partial}{\partial \tau} \int d^3 q\,  \psi^i_{\alpha,i}(\bq)\right)\,,
\end{align}
where we have used Eqs.~\re{eq:gradx}
and \re{eq:masscon2}. The same, obviously, holds for the integral of $\psi^i_{\beta,i}$.
Using the same procedure to get Eq.~\re{eq:lapeq}, Eq.~\re{eq:psia22} becomes
\begin{align}
&\frac{1}{2}\epsilon_{ikl}\epsilon_{jmn}(\delta^{km}+\psi^k_{\alpha,m})(\delta^{ln}+\psi^l_{\alpha,n})
 \frac{1}{a}\frac{\partial}{\partial \tau} \left(a \frac{\partial}{\partial \tau}\psi^i_{\alpha,j}\right)\nonumber\\
 &=\frac{3}{2}{\cal H}^2\Omega_m\left[\omega_\alpha \left(J_\alpha(\bq_\alpha)-1\right) + \omega_\beta \left(J_\alpha(\bq_\alpha)-J_{\beta\alpha}(\bq_\alpha)\right)\right]\,,
 \label{eq:lcdm2}
 \end{align}
 which contains precisely the $\bq_\alpha\to\bq_\beta$ jacobian, Eq.~\re{eq:detbetaalpha}.
 At second order, the source terms at the RHS contain the scalar contribution of Eq.~\re{eq:diff2}, giving 
 \beq
\psi^{(2)\,k}_\alpha = \cdots+ c^{(2),{\rm rel}}_{\beta\alpha}(\tau) \,\vp^{(2),{\rm rel}}_{\beta \alpha,k}\,,
\label{eq:psia2}
 \eeq
 and an analogous contribution to $\psi^{(2)\,k}_\beta$,
 \beq
 \psi^{(2)\,i}_\beta = \cdots+ c^{(2),{\rm rel}}_{\alpha\beta}(\tau) \,\vp^{(2),{\rm rel}}_{\alpha \beta,k}\,,
 \label{eq:psib}
 \eeq
 where the scalar $\vp^{(2),{\rm rel}}_{\alpha \beta}$ is obtained from Eq.~\re{eq:diff2} by exchanging $\alpha$ and $\beta$. 
 
 Momentum conservation gives a nontrivial constraint on the coefficients $c^{(2),{\rm rel}}_{\alpha\beta}(\tau)$ and $c^{(2),{\rm rel}}_{\beta\alpha}(\tau)$. Indeed, the contributions in Eqs.~\re{eq:psia2} and \re{eq:psib}, when integrated in $d^3 q$ do not vanish separately. On the other hand, considering their weighted avergage, 
 \begin{align}
 &\int d^3 q\,\left( \omega_\alpha \psi^{(2)\,k}_\alpha+ \omega_\beta \psi^{(2)\,k}_\beta\right)\,,\nonumber\\ & =\cdots + 
\int d^3 q \,\left(\omega_\alpha  c^{(2),{\rm rel}}_{\beta\alpha}(\tau) \,\vp^{(2),{\rm rel}}_{\beta \alpha,k} +\omega_\beta  c^{(2),{\rm rel}}_{\alpha\beta}(\tau) \,\vp^{(2),{\rm rel}}_{ \alpha\beta,k} \right) \,,\nonumber\\
&=\cdots + \int d^3q\, \frac{\partial_k \partial_i}{\nabla^2}\left[\left(\omega_\alpha  c^{(2),{\rm rel}}_{\beta\alpha} \vpl_{\beta,il} - \omega_\beta  c^{(2),{\rm rel}}_{\alpha\beta} \vpl_{\alpha,il}\right)\left(\vpl_{\alpha,l}- \vpl_{\beta,l}\right)\right]\,,
\end{align}
 the last integral vanishes upon integration by part  if 
 \beq
 \omega_\alpha \, c^{(2),{\rm rel}}_{\beta\alpha}(\tau)-\omega_\beta  \,c^{(2),{\rm rel}}_{\alpha\beta}(\tau)=0\,,
 \label{eq:momconst}
 \eeq
 enforcing momentum conservation as formulated in  Eq.~\re{eq:momcon2}.
 In $\Lambda$CDM the constraint is satisfied, as we can see from the equations of motion derived from Eq.~\re{eq:lcdm2},
 \begin{align}
\frac{1}{a D^2 }\frac{\partial}{\partial \tau} \left(a \frac{\partial}{\partial \tau}\left( c^{(2),{\rm rel}}_{\beta\alpha}(\tau) D^2 \right) \right) & = \frac{3}{2} \mathcal{H}^2 \Omega_m\,\omega_\beta,\nonumber\\
\frac{1}{a D^2 }\frac{\partial}{\partial \tau} \left(a \frac{\partial}{\partial \tau}\left( c^{(2),{\rm rel}}_{\alpha\beta}(\tau) D^2 \right) \right) & = \frac{3}{2} \mathcal{H}^2 \Omega_m\,\omega_\alpha,
 \end{align}
 which give $ c^{(2),{\rm rel}}_{\beta\alpha}\propto \omega_\beta$ and $ c^{(2),{\rm rel}}_{\alpha\beta}\propto \omega_\alpha$ (in the EdS limit we have $ c^{(2),{\rm rel}}_{\beta\alpha}/\omega_\beta =c^{(2),{\rm rel}}_{\alpha\beta}/\omega_\alpha=3/10$).
 
Therefore, at nonlinear orders,
in a generic cosmology, momentum conservation implies nontrivial relations between the
coefficients of terms like those in Eq.~\re{eq:momconst}, for the contributions to the displacements containing first derivatives.

In the multispecies case vorticity can be generated already at second order, unlike the single species case where it only emerges at third order, see Sect.~\ref{sect:n3}. Indeed, combining the two independent linear fields $\vpl_\alpha$ and $\vpl_\beta$ as in Eq.~\re{eq:transvect}, we can form a transverse vector,
\beq
v^i_{,kk}= \epsilon^{iln}\,\vpl_{\alpha,j l}\,\vpl_{\beta,j n}\,.
\label{eq:transvectmulti}
\eeq
This can be verified looking at the equations of motion for the vorticity of the two displacements,
\begin{align}
\frac{1}{a}\frac{\partial}{\partial \tau}\left(a \frac{\partial}{\partial \tau} \omega^{i}_{\alpha,mm}\right)
 &= -\epsilon^{iln} \psi^j_{\alpha, l} 
 \frac{1}{a}\frac{\partial}{\partial \tau}\left(a \frac{\partial}{\partial \tau} \psi^j_{\alpha,n}\right)
\,,
\end{align}
and the analogous one with $\alpha \to \beta$. Using the $\Lambda$CDM equation of motion, Eq.~\re{eq:lcdm2}, the time derivative at the RHS gives, already at linear order, a term proportional to $ \psi^j_{\beta,n}$, which generates a vorticity contribution of the form in Eq.~\re{eq:transvectmulti}.

\section{Conclusions}
\label{sec:conclusion}
An important task of ongoing and future galaxy surveys will be to look for deviations from the present Standard Model of Cosmology, the $\Lambda$CDM model. As an example, possible hints of a dynamical dark energy have already emerged in the Data Release 1 of \textsc{DESI} survey \cite{DESI:2024mwx}. Lacking compelling theoretical indications on a particular new physics scenario, the search should be as model-independent as possible. Moreover, the tightness  of already existing bounds, from CMB and LSS, tells us that the signatures, if any, will be tiny and their detection will require exploiting all possible observational handles in an optimal way. 

Present tests of $\Lambda$CDM consider effects at the background and the linear perturbation theory level. This work is a step towards model-independent tests in the non-linear sector of the theory, which is basically unconstrained so far, as most of the analyses consider the EdS approximation for the PT kernels (for analyses beyond this approximation in specific models, see \cite{DAmico:2020tty, Piga:2022mge, Amendola:2023awr}). We adopted the lagrangian formulation of PT, where the basic quantity is the displacement field. Although it is completely equivalent to the eulerian framework, it has some practical advantages, expecially if one is interested in field-level approaches to cosmological inference. For instance, the effect of large scale bulk flows is reproduced already at linear order in LPT, whereas it requires the summation of EPT contributions at all orders. 
More specifically to the bootstrap approach, imposing EGI implies an increasing number of constraints on the analytic structure of EPT kernels, whereas, in LPT, they are automatically satisfied by expressing the density contrast in terms of the displacement field. As a consequence, 
we were able to derive explicitly the  structure of the displacement field up to sixth order, and to outline an algorithmic procedure to go to arbitrarily higher orders.

The cosmology dependence is contained in the linear fields and in a set of time-dependent parameters which we computed explicitly in  $\Lambda$CDM, nDGP and wCDM models, but can also be left free, to parameterize the whole  class of theories sharing the same symmetries as $\Lambda$CDM. We also 
considered the case in which dark matter is composed of different non-relativistic species, showing that new structures, describing density and velocity biases, emerge.

The results of this work apply to dark matter fields, both in real and in redshift space. In future work, we will complete the construction of 
the LPT bootstrap forward model. On the one hand, the effect of short scale physics will be included by adding UV counterterms \cite{Baumann:2010tm, Pietroni:2011iz, Carrasco:2012cv, Porto:2013qua, Baldauf:2015tla}, on the other, biased tracers will be described in the same framework, by lifting the requirements of mass and momentum conservation, which leads to an increased number of admissible structures, parameterized by bias coefficients \cite{DAmico:2021rdb}.

\section*{Acknowledgements}
\noindent
We thank E. Castorina and G. D'Amico for useful discussions.
M. Marinucci and M. Pietroni acknowledge support by the MIUR Progetti di Ricerca di Rilevante Interesse Nazionale (PRIN) Bando 2022 - grant 20228RMX4A.
\appendix
\section{General structure of the displacement field beyond the fourth order}
\label{app:psi_beyond_neq4}
In this appendix, we will extend the results of Sect.~\ref{sect:PT} and perform an explicit calculation up to sixth order. We will also provide an algorithm to generalize the results to arbitrary $n$-th order.

\subsection{Case $n=5$}
At fifth order, we can build three scalars of the form Eq.~\re{eq:nl3},
\begin{align}
\vp^{(5)}_{1,ii} &\equiv \frac{1}{3!} \epsilon^{i_1 i_2 i_3}\,\epsilon^{j_1 j_2 j_3}\, \vp^{(3)}_{1,i_1 j_1}\vp^{(1)}_{,i_2 j_2}\vp^{(1)}_{,i_3 j_3}\,\nonumber\\
\vp^{(5)}_{2,ii} &\equiv \frac{1}{3!} \epsilon^{i_1 i_2 i_3}\,\epsilon^{j_1 j_2 j_3}\, \vp^{(3)}_{2,i_1 j_1}\vp^{(1)}_{,i_2 j_2}\vp^{(1)}_{,i_3 j_3}\,\nonumber\\
\vp^{(5)}_{3,ii} &\equiv \frac{1}{3!} \epsilon^{i_1 i_2 i_3}\,\epsilon^{j_1 j_2 j_3}\, \vp^{(2)}_{,i_1 j_1}\vp^{(2)}_{,i_2 j_2}\vp^{(1)}_{,i_3 j_3}\,,
\end{align}
and seven of the form Eq.~\re{eq:nl2},
\begin{align}
\vp^{(5)}_{4,ii} &\equiv \frac{1}{2!} \epsilon^{i_1 i_2 k}\,\epsilon^{j_1 j_2 k}\, \vp^{(3)}_{1,i_1 j_1}\vp^{(2)}_{,i_2 j_2}\,\nonumber\\
\vp^{(5)}_{5,ii} &\equiv \frac{1}{2!} \epsilon^{i_1 i_2 k}\,\epsilon^{j_1 j_2 k}\, \vp^{(3)}_{2,i_1 j_1}\vp^{(2)}_{,i_2 j_2}\,\nonumber\\
\vp^{(5)}_{6,ii} &\equiv \frac{1}{2!} \epsilon^{i_1 i_2 k}\,\epsilon^{j_1 j_2 k}\, \vp^{(4)}_{1,i_1 j_1}\vp^{(1)}_{,i_2 j_2}\,\nonumber\\
\vp^{(5)}_{7,ii} &\equiv \frac{1}{2!} \epsilon^{i_1 i_2 k}\,\epsilon^{j_1 j_2 k}\, \vp^{(4)}_{2,i_1 j_1}\vp^{(1)}_{,i_2 j_2}\,\nonumber\\
\vp^{(5)}_{8,ii} &\equiv \frac{1}{2!} \epsilon^{i_1 i_2 k}\,\epsilon^{j_1 j_2 k}\, \vp^{(4)}_{3,i_1 j_1}\vp^{(1)}_{,i_2 j_2}\,\nonumber\\
\vp^{(5)}_{9,ii} &\equiv \frac{1}{2!} \epsilon^{i_1 i_2 k}\,\epsilon^{j_1 j_2 k}\, \vp^{(4)}_{4,i_1 j_1}\vp^{(1)}_{,i_2 j_2}\,\nonumber\\
\vp^{(5)}_{10,ii} &\equiv \frac{1}{2!} \epsilon^{i_1 i_2 k}\,\epsilon^{j_1 j_2 k}\, \vp^{(4)}_{5,i_1 j_1}\vp^{(1)}_{,i_2 j_2}\,.
\end{align}

As in the previous case with $n=4$, one can also generate scalar terms using the lower order vector fields,
\begin{align}
\vp^{(5)}_{11,ii} &=\frac{1}{2}\vp^{(2)}_{,lj}\left(\epsilon^{jmn} v^{(3)\,n}_{,lm}+ \epsilon^{lmn} v^{(3)\,n}_{,jm} \right)\,\nonumber\\
\vp^{(5)}_{12,ii} &=\frac{1}{2}\vp^{(1)}_{,lj}\left(\epsilon^{jmn} v^{(4)\,n}_{1,lm}+ \epsilon^{lmn} v^{(4)\,n}_{1,jm} \right)\,\nonumber\\
\vp^{(5)}_{13,ii} &=\frac{1}{2}\vp^{(1)}_{,lj}\left(\epsilon^{jmn} v^{(4)\,n}_{2,lm}+ \epsilon^{lmn} v^{(4)\,n}_{2,jm} \right)\,\nonumber\\
\vp^{(5)}_{14,ii} &=\frac{1}{2}\vp^{(1)}_{,lj}\left(\epsilon^{jmn} v^{(4)\,n}_{3,lm}+ \epsilon^{lmn} v^{(4)\,n}_{3,jm} \right).
\end{align}
Starting at this order, we also have a contribution of the form
\beq
\vp^{(5)}_{15,ii} = \frac{1}{3!}\epsilon^{ikl}\epsilon^{jmn} \epsilon^{lop} \vpl_{,ij}\vpl_{,km} v^{(3)\,p}_{,on},
\eeq
which can be verified to be among the terms that respect mass conservation (see Eq.~\ref{eq:masscon2}).

In summary, the most generic scalar component of the displacement field at fifth order contains 15 independent contributions,
\beq
\phi^{(5)}(\bq,\tau)=\sum_{a=1}^{15} c^{(5)}_a(\tau) \vp^{(5)}_a(\bq,\tau)\,.
\eeq

As for the vectors, following the general structure Eq.~\re{eq:nlvort}, we get 
\begin{align}
v_{1,kk}^{(5)\, i} &\equiv  \, \epsilon^{iln} \,\varphi^{(4)}_{1,jl} \, \varphi^{(1)}_{,jn} \,\nonumber\\
v_{2,kk}^{(5)\, i} &\equiv  \, \epsilon^{iln} \,\varphi^{(4)}_{2,jl} \, \varphi^{(1)}_{,jn} \,\nonumber\\
v_{3,kk}^{(5)\, i} &\equiv  \, \epsilon^{iln} \,\varphi^{(4)}_{3,jl} \, \varphi^{(1)}_{,jn} \,\nonumber\\
v_{4,kk}^{(5)\, i} &\equiv  \, \epsilon^{iln} \,\varphi^{(4)}_{4,jl} \, \varphi^{(1)}_{,jn} \,\nonumber\\
v_{5,kk}^{(5)\, i} &\equiv  \, \epsilon^{iln} \,\varphi^{(4)}_{5,jl} \, \varphi^{(1)}_{,jn} \,\nonumber\\
v_{6,kk}^{(5)\, i} &\equiv  \, \epsilon^{iln} \,\varphi^{(3)}_{1,jl} \, \varphi^{(2)}_{,jn} \,\nonumber\\
v_{7,kk}^{(5)\, i} &\equiv  \, \epsilon^{iln} \,\varphi^{(3)}_{2,jl} \, \varphi^{(2)}_{,jn} \,\nonumber\\
v_{8,kk}^{(5)\, i} &\equiv \epsilon^{iln}\,\epsilon^{jmp}\vpl_{,jl} v^{(4)\,p}_{1,mn}\,\nonumber\\
v_{9,kk}^{(5)\, i} &\equiv \epsilon^{iln}\,\epsilon^{jmp}\vpl_{,jl} v^{(4)\,p}_{2,mn}\,\nonumber\\
v_{10,kk}^{(5)\, i} &\equiv \epsilon^{iln}\,\epsilon^{jmp}\vpl_{,jl} v^{(4)\,p}_{3,mn}\,\nonumber\\
v_{11,kk}^{(5)\, i} &\equiv \epsilon^{iln}\,\epsilon^{jmp}\vp^{(2)}_{,jl} v^{(3)\,p}_{,mn}\,.
\end{align}
Therefore the most generic vector contribution to the displacement field at $n=5$ carries 11 cosmology dependent coefficients,
\beq
\omega^{(5)\, i}(\bq, \tau) =\sum_{a=1}^{11} d^{(5)}_a(\tau) v_{a}^{(5)\, i}(\bq, \tau) \,.
\eeq

\subsection{Case $n=6$}
In the case of sixth order, we can build 8 scalars of the form Eq.~\re{eq:nl3},
\begin{align}
\vp^{(6)}_{1,ii} &\equiv \frac{1}{3!} \epsilon^{i_1 i_2 i_3}\,\epsilon^{j_1 j_2 j_3}\, \vp^{(4)}_{1,i_1 j_1}\vp^{(1)}_{,i_2 j_2}\vp^{(1)}_{,i_3 j_3}\,\nonumber\\
\vp^{(6)}_{2,ii} &\equiv \frac{1}{3!} \epsilon^{i_1 i_2 i_3}\,\epsilon^{j_1 j_2 j_3}\, \vp^{(4)}_{2,i_1 j_1}\vp^{(1)}_{,i_2 j_2}\vp^{(1)}_{,i_3 j_3}\,\nonumber\\
\vp^{(6)}_{3,ii} &\equiv \frac{1}{3!} \epsilon^{i_1 i_2 i_3}\,\epsilon^{j_1 j_2 j_3}\, \vp^{(4)}_{3,i_1 j_1}\vp^{(1)}_{,i_2 j_2}\vp^{(1)}_{,i_3 j_3}\,\nonumber\\
\vp^{(6)}_{4,ii} &\equiv \frac{1}{3!} \epsilon^{i_1 i_2 i_3}\,\epsilon^{j_1 j_2 j_3}\, \vp^{(4)}_{4,i_1 j_1}\vp^{(1)}_{,i_2 j_2}\vp^{(1)}_{,i_3 j_3}\,\nonumber\\
\vp^{(6)}_{5,ii} &\equiv \frac{1}{3!} \epsilon^{i_1 i_2 i_3}\,\epsilon^{j_1 j_2 j_3}\, \vp^{(4)}_{5,i_1 j_1}\vp^{(1)}_{,i_2 j_2}\vp^{(1)}_{,i_3 j_3}\,\nonumber\\
\vp^{(6)}_{6,ii} &\equiv \frac{1}{3!} \epsilon^{i_1 i_2 i_3}\,\epsilon^{j_1 j_2 j_3}\, \vp^{(3)}_{1,i_1 j_1}\vp^{(2)}_{,i_2 j_2}\vp^{(1)}_{,i_3 j_3}\,\nonumber\\
\vp^{(6)}_{7,ii} &\equiv \frac{1}{3!} \epsilon^{i_1 i_2 i_3}\,\epsilon^{j_1 j_2 j_3}\, \vp^{(3)}_{2,i_1 j_1}\vp^{(2)}_{,i_2 j_2}\vp^{(1)}_{,i_3 j_3}\,\nonumber\\
\vp^{(6)}_{8,ii} &\equiv \frac{1}{3!} \epsilon^{i_1 i_2 i_3}\,\epsilon^{j_1 j_2 j_3}\, \vp^{(2)}_{,i_1 j_1}\vp^{(2)}_{,i_2 j_2}\vp^{(2)}_{,i_3 j_3}\,,
\end{align}
and 23 of the form Eq.~\re{eq:nl2},
\begin{align}
\vp^{(6)}_{9,ii} &\equiv \frac{1}{2!} \epsilon^{i_1 i_2 k}\,\epsilon^{j_1 j_2 k}\, \vp^{(4)}_{1,i_1 j_1}\vp^{(2)}_{,i_2 j_2}\,\nonumber\\
\vp^{(6)}_{10,ii} &\equiv \frac{1}{2!} \epsilon^{i_1 i_2 k}\,\epsilon^{j_1 j_2 k}\, \vp^{(4)}_{2,i_1 j_1}\vp^{(2)}_{,i_2 j_2}\,\nonumber\\
\vp^{(6)}_{11,ii} &\equiv \frac{1}{2!} \epsilon^{i_1 i_2 k}\,\epsilon^{j_1 j_2 k}\, \vp^{(4)}_{3,i_1 j_1}\vp^{(2)}_{,i_2 j_2}\,\nonumber\\
\vp^{(6)}_{12,ii} &\equiv \frac{1}{2!} \epsilon^{i_1 i_2 k}\,\epsilon^{j_1 j_2 k}\, \vp^{(4)}_{4,i_1 j_1}\vp^{(2)}_{,i_2 j_2}\,\nonumber\\
\vp^{(6)}_{13,ii} &\equiv \frac{1}{2!} \epsilon^{i_1 i_2 k}\,\epsilon^{j_1 j_2 k}\, \vp^{(4)}_{5,i_1 j_1}\vp^{(2)}_{,i_2 j_2}\,\nonumber\\
\vp^{(6)}_{14,ii} &\equiv \frac{1}{2!} \epsilon^{i_1 i_2 k}\,\epsilon^{j_1 j_2 k}\, \vp^{(3)}_{1,i_1 j_1}\vp^{(3)}_{1,i_2 j_2}\,\nonumber\\
\vp^{(6)}_{15,ii} &\equiv \frac{1}{2!} \epsilon^{i_1 i_2 k}\,\epsilon^{j_1 j_2 k}\, \vp^{(3)}_{2,i_1 j_1}\vp^{(3)}_{1,i_2 j_2}\,\nonumber\\
\vp^{(6)}_{16,ii} &\equiv \frac{1}{2!} \epsilon^{i_1 i_2 k}\,\epsilon^{j_1 j_2 k}\, \vp^{(3)}_{2,i_1 j_1}\vp^{(3)}_{2,i_2 j_2}\,\nonumber\\
\vp^{(6)}_{17,ii} &\equiv \frac{1}{2!} \epsilon^{i_1 i_2 k}\,\epsilon^{j_1 j_2 k}\, \vp^{(5)}_{1,i_1 j_1}\vp^{(1)}_{,i_2 j_2}\,\nonumber\\
&\vdots \,\nonumber\\
\vp^{(6)}_{31,ii} &\equiv \frac{1}{2!} \epsilon^{i_1 i_2 k}\,\epsilon^{j_1 j_2 k}\, \vp^{(5)}_{15,i_1 j_1}\vp^{(1)}_{,i_2 j_2}\,.
\end{align}

Using the lower order vector fields, one can also generate terms of the following form
\begin{align}
\vp^{(6)}_{32,ii} &=\frac{1}{2}\vp^{(3)}_{1,lj}\left(\epsilon^{jmn} v^{(3)\,n}_{,lm}+ \epsilon^{lmn} v^{(3)\,n}_{,jm} \right)\,\nonumber\\
\vp^{(6)}_{33,ii} &=\frac{1}{2}\vp^{(3)}_{2,lj}\left(\epsilon^{jmn} v^{(3)\,n}_{,lm}+ \epsilon^{lmn} v^{(3)\,n}_{,jm} \right)\,\nonumber\\
\vp^{(6)}_{34,ii} &=\frac{1}{2}\vp^{(2)}_{,lj}\left(\epsilon^{jmn} v^{(4)\,n}_{1,lm}+ \epsilon^{lmn} v^{(4)\,n}_{1,jm} \right)\,\nonumber\\
\vp^{(6)}_{35,ii} &=\frac{1}{2}\vp^{(2)}_{,lj}\left(\epsilon^{jmn} v^{(4)\,n}_{2,lm}+ \epsilon^{lmn} v^{(4)\,n}_{2,jm} \right)\,\nonumber\\
\vp^{(6)}_{36,ii} &=\frac{1}{2}\vp^{(2)}_{,lj}\left(\epsilon^{jmn} v^{(4)\,n}_{3,lm}+ \epsilon^{lmn} v^{(4)\,n}_{3,jm} \right)\,\nonumber\\
\vp^{(6)}_{37,ii} &=\frac{1}{2}\vp^{(1)}_{,lj}\left(\epsilon^{jmn} v^{(5)\,n}_{1,lm}+ \epsilon^{lmn} v^{(5)\,n}_{1,jm} \right)\,\nonumber\\
&\vdots \nonumber \\
\vp^{(6)}_{47,ii} &=\frac{1}{2}\vp^{(1)}_{,lj}\left(\epsilon^{jmn} v^{(5)\,n}_{11,lm}+ \epsilon^{lmn} v^{(5)\,n}_{11,jm} \right)\,\nonumber\\
\end{align}
or of the following form
\begin{align}
\vp^{(6)}_{48,ii} &= \frac{1}{3!} \epsilon^{ikl}\epsilon^{jmn} v^{(3)\,l}_{,jk}\,v^{(3)\,n}_{,im}
\end{align}
and of the following form
\begin{align}
\vp^{(6)}_{49,ii} &= \frac{1}{3!}\epsilon^{ikl}\epsilon^{jmn} \epsilon^{lop} \vpl_{,ij}\vpl_{,km} v^{(4)\,p}_{1,on}\,\nonumber\\
\vp^{(6)}_{50,ii} &= \frac{1}{3!}\epsilon^{ikl}\epsilon^{jmn} \epsilon^{lop} \vpl_{,ij}\vpl_{,km} v^{(4)\,p}_{2,on}\,\nonumber\\
\vp^{(6)}_{51,ii} &= \frac{1}{3!}\epsilon^{ikl}\epsilon^{jmn} \epsilon^{lop} \vpl_{,ij}\vpl_{,km} v^{(4)\,p}_{3,on}\,\nonumber\\
\vp^{(6)}_{52,ii} &= \frac{1}{3!}\epsilon^{ikl}\epsilon^{jmn} \epsilon^{lop} \vpl_{,ij}\vp^{(2)}_{,km} v^{(3)\,p}_{,on}\,.
\end{align}
In summary, we have 52 independent contributions for the most generic scalar component of the displacement field at sixth order,
\beq
\phi^{(6)}(\bq,\tau)=\sum_{a=1}^{52} c^{(6)}_a(\tau) \vp^{(6)}_a(\bq,\tau)\,.
\eeq

Following the general structure Eq.~\re{eq:nlvort}, vectors contribution to the displacement field at sixth order are as follows 
\begin{align}
v_{1,kk}^{(6)\, i} &\equiv  \, \epsilon^{iln} \,\varphi^{(5)}_{1,jl}  \varphi^{(1)}_{,jn} \,\nonumber\\
&\vdots \nonumber \\
v_{15,kk}^{(6)\, i} &\equiv  \, \epsilon^{iln} \,\varphi^{(5)}_{15,jl}  \varphi^{(1)}_{,jn} \,\nonumber\\
v_{16,kk}^{(6)\, i} &\equiv  \, \epsilon^{iln} \,\varphi^{(4)}_{1,jl}  \varphi^{(2)}_{,jn} \,\nonumber\\
v_{17,kk}^{(6)\, i} &\equiv  \, \epsilon^{iln} \,\varphi^{(4)}_{2,jl}  \varphi^{(2)}_{,jn} \,\nonumber\\
v_{18,kk}^{(6)\ i} &\equiv  \, \epsilon^{iln} \,\varphi^{(4)}_{3,jl}  \varphi^{(2)}_{,jn} \,\nonumber\\
v_{19,kk}^{(6)\, i} &\equiv  \, \epsilon^{iln} \,\varphi^{(4)}_{4,jl}  \varphi^{(2)}_{,jn} \,\nonumber\\
v_{20,kk}^{(6)\, i} &\equiv  \, \epsilon^{iln} \,\varphi^{(4)}_{5,jl}  \varphi^{(2)}_{,jn} \,\nonumber\\
v_{21,kk}^{(6)\, i} &\equiv \epsilon^{iln}\,\epsilon^{jmp}\vpl_{,jl} v^{(5)\,p}_{1,mn}\,\nonumber\\
&\vdots \nonumber \\
v_{31,kk}^{(6)\, i} &\equiv \epsilon^{iln}\,\epsilon^{jmp}\vpl_{,jl} v^{(5)\,p}_{11,mn}\,\nonumber\\
v_{32,kk}^{(6)\, i} &\equiv \epsilon^{iln}\,\epsilon^{jmp}\vp^{(2)}_{,jl} v^{(4)\,p}_{1,mn}\,\nonumber\\
v_{33,kk}^{(6)\, i} &\equiv \epsilon^{iln}\,\epsilon^{jmp}\vp^{(2)}_{,jl} v^{(4)\,p}_{2,mn}\,\nonumber\\
v_{34,kk}^{(6)\, i} &\equiv \epsilon^{iln}\,\epsilon^{jmp}\vp^{(2)}_{,jl} v^{(4)\,p}_{3,mn}\,\nonumber\\
v_{35,kk}^{(6)\, i} &\equiv \epsilon^{iln}\,\epsilon^{jrs}\,\epsilon^{jmp} v^{(3)\,r}_{,sl} v^{(3)\,p}_{,mn}\,
\end{align}
Therefore the most generic vector carries 35 cosmology dependent coefficients,
\beq
\omega^{(6)\, i}(\bq, \tau) =\sum_{a=1}^{35} d^{(6)}_a(\tau) v_{a}^{(6)\, i}(\bq, \tau) \,.
\eeq

\subsection{The general case of the $n$-th order}

In this section, we describe an algorithm to generate the scalar and vector contributions to the displacement field at the $n$-th order. We start with the linear scalar field $\varphi^{(1)}$.  First, we would like to recursively build the set of operators $\mathcal{O}_\varphi^{(m)}$ which contain the higher-order scalar contributions to the displacement field $\varphi^{(m)} \in \mathcal{O}_\varphi^{(m)}$ for $m = 2, \cdots, n$. Generally, following Eqs.~\eqref{eq:nl3} and \eqref{eq:nl2}, we can see that the general scalar field can be constructed out of the product of the scalar and vector fields of the lower order. In other words
\begin{equation}
    \mathcal{O}_\varphi^{(m)} \equiv \mathcal{O}^{(m)}_{\varphi, SS} ~\cup~ \mathcal{O}^{(m)}_{\varphi, SV} ~\cup~ \mathcal{O}^{(m)}_{\varphi, VV} ~\cup~ \mathcal{O}^{(m)}_{\varphi, SSS} ~\cup~ \mathcal{O}^{(m)}_{\varphi, SSV} ~\cup~\mathcal{O}^{(m)}_{\varphi, SVV} ~\cup~\mathcal{O}^{(m)}_{\varphi, VVV},
\end{equation}
where we have denoted the operators by the subscript $S$ and $V$ for scalar and (transverse) vector respectively and defined the following set of operators
\begin{align}
\mathcal{O}^{(m)}_{\varphi, SS} & \equiv \Bigg \{ \frac{1}{2}\left(\varphi^{(m_1)}_{a_1,ii}\varphi^{(m_2)}_{a_2,jj}-\varphi^{(m_2)}_{a_1,ij}\varphi^{(m_2)}_{a_2,ji}\right) ~\Bigg |~ m_1 + m_2 = m, a_i =  1,\cdots, |\mathcal{O}_\varphi^{(m_i)}| \Bigg \},
\\
\mathcal{O}^{(m)}_{\varphi, SV} &\equiv \Bigg \{\epsilon^{jmn} \varphi^{(m_1)}_{a_1,ij} v^{(m_2)\, n}_{b_2,im} ~\Bigg|~ m_1 + m_2 = m, a_i =  1,\cdots, |\mathcal{O}_\varphi^{(m_i)}|, b_i =  1,\cdots, |\mathcal{O}_v^{(m_i)}| \Bigg \},
\\
\mathcal{O}^{(m)}_{\varphi, VV} &\equiv \Bigg \{\epsilon^{ikl}\epsilon^{jmn} v^{(m_1)\,l}_{b_1,jk}\, v^{(m_2)\,n}_{b_2,im} ~\Bigg|~ m_1 + m_2 = m, b_i =  1,\cdots, |\mathcal{O}_v^{(m_i)}| \Bigg \},
\\
\mathcal{O}^{(m)}_{\varphi, SSS} &\equiv \Bigg \{\frac{1}{3!}\epsilon^{ikl}\epsilon^{jmn}\varphi^{(m_1)}_{a_1,ij}\varphi^{(m_2)}_{a_2,km}\varphi^{(m_3)}_{a_3,ln} ~\Bigg|~ m_1 + m_2 + m_3 = m, a_i =  1,\cdots, |\mathcal{O}_\varphi^{(m_i)}| \Bigg \},
\\
\mathcal{O}^{(m)}_{\varphi, SSV} &\equiv \Bigg \{\frac{1}{2}\epsilon^{jmn} \varphi^{(m_1)}_{a_1,ij} \varphi^{(m_2)}_{a_2,km} \left( v_{b_3,in}^{(m_3)\,k} - v_{a_3,kn}^{(m_3)\,i} \right) ~\Bigg|~ m_1 + m_2 + m_3 = m,
\nonumber \\& \hspace{3 cm}
a_i =  1,\cdots, |\mathcal{O}_\varphi^{(m_i)}|,  b_i =  1,\cdots, |\mathcal{O}_v^{(m_i)}| \Bigg \},
\\
\mathcal{O}^{(m)}_{\varphi, SVV} &\equiv \Bigg \{\frac{1}{2} \epsilon^{jmn} \epsilon^{lop} \varphi^{(m_1)}_{a_1,ij} v^{(m_2)\, p}_{b_2,on} \left( v^{(m_3)\, i}_{b_3,lm} - v^{(m_3)\, l}_{b_3,im} \right) ~\Bigg|~ m_1 + m_2 + m_3 = m, 
\nonumber \\& \hspace{3 cm}
a_i =  1,\cdots, |\mathcal{O}_\varphi^{(m_i)}|, b_i =  1,\cdots, |\mathcal{O}_v^{(m_i)}| \Bigg \},
\\
\mathcal{O}^{(m)}_{\varphi, VVV} &\equiv \Bigg \{\frac{1}{3!} \epsilon^{jmn} \epsilon^{kqr} \epsilon^{lop} v^{(m_1)\, r}_{b_1,qm} v^{(m_2)\, p}_{b_2,on} \left( v^{(m_3)\, l}_{b_3,jk} - v^{(m_3)\, k}_{b_3,jl} \right) ~\Bigg|~ m_1 + m_2 + m_3 = m, b_i =  1,\cdots, |\mathcal{O}_v^{(m_i)}| \Bigg \},
\end{align}
where $|\mathcal{O}_\vp^{(m_i)}|$ and $|\mathcal{O}_v^{(m_i)}|$ represent the number of elements in the sets $\mathcal{O}_\vp^{(m_i)}$ and $\mathcal{O}_v^{(m_i)}$, respectively. 

We note that the above definitions also depend on the similar set of operators $\mathcal{O}^{(m)}_v$ that contain the higher-order (transverse) vectors $v^{(m)\, i} \in \mathcal{O}^{(m)}_v$ for $m = 2, \cdots, n$. These operators, following Eq.~\eqref{eq:nlv}, are constructed from the following set of operators
\begin{equation}
    \mathcal{O}^{(m)}_v = \mathcal{O}^{(m)}_{v, SS} ~ \cup ~ \mathcal{O}^{(m)}_{v, SV} ~ \cup ~ \mathcal{O}^{(m)}_{v, VV},
\end{equation}
with
\begin{align}
\mathcal{O}^{(m)}_{v, SS} &\equiv \Bigg \{ \epsilon^{iln} \varphi^{(m_1)}_{a_1,lj} \varphi^{(m_2)}_{a_2,ln} ~\Bigg |~ m_1 + m_2 = m, a_i = 1, \cdots, |\mathcal{O}_\varphi^{(m_i)}| \Bigg \},
\\
\mathcal{O}^{(m)}_{v, SV} &\equiv \Bigg \{ \epsilon^{iln} \epsilon^{jpq} \varphi^{(m_1)}_{a_1,lj} v^{(m_2)\,q}_{b_2,pn} ~\Bigg |~ m_1 + m_2 = m, a_i = 1, \cdots, |\mathcal{O}_\varphi^{(m_i)}|, b_i = 1, \cdots, |\mathcal{O}_v^{(m_i)}|\Bigg \},
\\
\mathcal{O}^{(m)}_{v, VV} &\equiv \Bigg \{ \epsilon^{iln} v^{(m_1)\, q}_{b_1,pn} \left( v^{(m_2)\, q}_{b_2,pl} - 
v^{(m_2)\, p}_{b_2,ql} \right) ~\Bigg |~ m_1 + m_2 = m, b_i = 1, \cdots, |\mathcal{O}_v^{(m_i)}| \Bigg \}.
\end{align}

Therefore, at any given order $n$, we can build the scalars
\begin{equation}
\phi^{(n)} (\bq, \tau) = \sum_{a = 1, \cdots, | \mathcal{O}_\varphi^{(n)}|}  c_a(\tau) \varphi_a^{(n)}(\bq, \tau)
\end{equation}
and the (transverse) vectors
\begin{equation}
\omega^{(n), i} (\bq, \tau) = \sum_{a = 1, \cdots, | \mathcal{O}_v^{(n)}|}  d_a(\tau) v_a^{(n)\, i}(\bq, \tau),
\end{equation}
which, in turn, construct the displacement field at the $n$-th order
\begin{equation}
    \psi^{(n)\, i} (\bq, \tau)= \partial_i \phi^{(n)} (\bq, \tau)+ \epsilon^{ijk} \partial_j \omega^{(n)\, k} (\bq, \tau).
\end{equation}

\section{Vorticity at fourth order}
\label{app:AppVort}
At fourth order, the equation of motion for vorticity, Eq.~\re{eq:vorttot}, gives
\begin{align}
\frac{1}{a}\frac{\partial}{\partial \tau} \left(a \frac{\partial}{\partial \tau} \omega^{(4),i}_{,mm}  \right) = &- \epsilon^{ijk} \left[ \varphi^{(3)}_{1,lj} \, \frac{1}{a}\frac{\partial}{\partial \tau} \left(a \frac{\partial}{\partial \tau} \varphi^{(1)}_{,lk} \right) -  \frac{1}{a}\frac{\partial}{\partial \tau} \left(a \frac{\partial}{\partial \tau} \varphi^{(3)}_{1,lj} \right) \, \varphi^{(1)}_{,lk}
\right]
\nonumber\\ &
- \epsilon^{ijk} \left[ \varphi^{(3)}_{2,lj} \, \frac{1}{a}\frac{\partial}{\partial \tau} \left(a \frac{\partial}{\partial \tau} \varphi^{(1)}_{,lk} \right) -  \frac{1}{a}\frac{\partial}{\partial \tau} \left(a \frac{\partial}{\partial \tau} \varphi^{(3)}_{2,lj} \right) \, \varphi^{(1)}_{,lk}
\right]
\nonumber\\ &
- \epsilon^{ijk} \epsilon^{lpq} \left[ \omega^{(3),q}_{,pj} \, \frac{1}{a}\frac{\partial}{\partial \tau} \left(a \frac{\partial}{\partial \tau} \varphi^{(1)}_{,lk} \right)  - \frac{1}{a}\frac{\partial}{\partial \tau} \left(a \frac{\partial}{\partial \tau} \omega^{(3),q}_{,pj} \right) \, \varphi^{(1)}_{,lk}
\right].
\label{eq:vort4th}
\end{align}
The general solution at fourth order is given in Eq.~\re{eq:vv4}, which, inserted in \re{eq:vort4th}, leads to the following equations for the coefficients,
\begin{align}
\frac{1}{a D^4}\frac{\partial}{\partial \tau} \left(a \frac{\partial}{\partial \tau} \left(d_1^{(4)}(\tau) D^4 \right)  \right) &= -3 \mathcal{H}^2 \Omega_m
\,,\nonumber \\
\frac{1}{a D^4}\frac{\partial}{\partial \tau} \left(a \frac{\partial}{\partial \tau} \left(d_2^{(4)}(\tau) D^4 \right)  \right) &= -3 \mathcal{H}^2 \Omega_m \left( 1-c^{(2)} \right)
\,,\nonumber \\
\frac{1}{a D^4}\frac{\partial}{\partial \tau} \left(a \frac{\partial}{\partial \tau} \left(d_3^{(4)}(\tau) D^4 \right)  \right) &= - \frac{3}{2}\mathcal{H}^2\Omega_m \left(d^{(3)} - 1\right) .
\end{align}
In the EdS limit we have $\{d_1^{(4)}(\tau), d_2^{(4)}(\tau), d_3^{(4)}(\tau)\} = \{-1/6, -5/21, 1/14\}$.
\section{Equations for modified gravity theories}
\label{app:MG}

We report here the time evolution equations for the third order coefficients $c_1^{(3)}$ and $c_2^{(3)}$ in a modified gravity scenario, nDGP. They are
\begin{align}
    \frac{d^2 c_1^{(3)}}{d\chi^2} + \frac{d c_1^{(3)}}{d\chi} \left(5 + \frac{3}{2}\frac{\Omega_m}{f^2}\mu \right) + 2 c_1^{(3)}\left(3 + \frac{3}{2}\frac{\Omega_m}{f^2} \mu\right) =& - 3 \frac{dc^{(2)}}{d\chi}\left(\frac{3}{2}\frac{\Omega_m}{f^2} 
    - \frac{3}{2}\frac{\Omega_m}{f^2}\mu\right)\nonumber\\
    &- 3 \frac{\Omega_m}{f^2} - 6 \left(\frac{3}{2}\Omega_m\right)^2\frac{\mu_2}{f^2}\,,
\end{align}
\begin{align}
    \frac{d^2 c_2^{(3)}}{d\chi^2} + \frac{d c_2^{(3)}}{d\chi} \left(5 + \frac{3}{2}\frac{\Omega_m}{f^2}\mu \right) + 2 c_2^{(3)}\left(3 + \frac{3}{2}\frac{\Omega_m}{f^2} \mu\right) =& 2 c^{(2)}\left(\frac{3}{2}\frac{\Omega_m}{f^2} + 2 \left(\frac{3}{2}\Omega_m\right)^2\frac{\mu_2}{f^2}\right)\nonumber\\
    &- 2 \frac{dc^{(2)}}{d\chi}\left(\frac{3}{2}\frac{\Omega_m}{f^2} - \frac{3}{2}\frac{\Omega_m}{f^2}\mu\right)\nonumber\\
    &- 3 \frac{\Omega_m}{f^2}\mu -4\left(\frac{3}{2}\Omega_m\right)^2\left(\frac{\mu_2}{f^2} + \frac{3}{2}\frac{\Omega_m}{f^2}\mu_{22}\right)\,,
\end{align}
where the $\mu$'s function read
\begin{align}
    \mu = 1 + \frac{1}{3\beta}\,,\quad\mu_2 = -\frac{1}{2}\left(\frac{H}{H_0}\right)^2\frac{1}{\Omega_{\rm rc}} \left(\frac{1}{3\beta}\right)^3\,\quad\mu_{22} = 2 \left(\frac{H}{H_0}\right)^4\frac{1}{\Omega_{\rm rc}^2}\left(\frac{1}{3\beta}\right)^2\,,
\end{align}
with 
\begin{equation}
    \beta = 1 + \frac{H}{H_0}\frac{1}{\sqrt{\Omega_{\rm rc}}}\left(1 + \frac{1}{3}\frac{d\log{H}}{d\log{a}}\right)\,.
\end{equation}
$\Lambda$CDM is recovered when $\Omega_{\rm rc}\to 0$.
\section{Mapping between LPT and EPT bootstrap}
\label{app:EPTmap}
In this section we derive the LPT bootstrap parameters introduced in this paper and the EPT ones of \cite{DAmico:2021rdb}. We use here the notation of \cite{Amendola:2023awr} for the EPT kernels.
We start by defining the functions 
\begin{align}
\alpha_s(\bq_1,\bq_2) & =1+\frac{\bq_{1}\cdot \bq_{2}}{2}\left(\frac{1}{q_{1}^{2}}+\frac{1}{q_{2}^{2}}\right)\\
\beta(\bq_{1},\bq_{2}) & =\frac{|\bq_{1}+\bq_{2}|^{2}\bq_{1}\cdot \bq_{2}}{2q_{1}^{2}q_{2}^{2}}\\
\gamma(\bq_{1},\bq_{2}) & =1-\frac{(\bq_{1}\cdot \bq_{2})^{2}}{q_{1}^{2}q_{2}^{2}}=\alpha_{s}(\bq_{1},\bq_{2})-\beta(\bq_{1},\bq_{2})\\
\alpha_{a}(\bq_{1},\bq_{2}) & =\frac{\bq_{1}\cdot \bq_{2}}{q_{1}^{2}}-\frac{\bq_{2}\cdot \bq_{1}}{q_{2}^{2}}
\end{align}
The matter kernels are then given by,
\begin{align}
& 2\,F_{2}(\bq_1,\bq_2; z)   =2\, \beta(\bq_{1},\bq_{2})+ a_{\gamma}^{(2)}(z)\,\gamma(\bq_{1},\bq_{2})\,,\label{eq:F2}\\
& 3!\, F_{3}(\bq_1,\bq_2,\bq_3; z)  =2\, O_{\beta\beta}(\bq_{1},\bq_{2},\bq_{3})
+\bigg[a_{\gamma a}^{(3)}(z)\nonumber \\
& \quad -2\left(a_{\gamma b}^{(3)}(z) - 1 \right)\bigg]O_{\beta\gamma}(\bq_{1},\bq_{2},\bq_{3}) +\left(\frac{1}{4} a_{\gamma a}^{(3)}(z)-\frac{1}{2}a_{\gamma b}^{(3)}(z)\right)O_{\gamma\alpha_{a}}(\bq_{1},\bq_{2},\bq_{3})\nonumber\\
 & \quad +\left(a_{\gamma b}^{(3)}(z)-\frac{1}{2}a_{\gamma a}^{(3)}(z)+2(a_{\gamma}^{(2)}(z) - 1)\right) O_{\gamma\beta}(\bq_{1},\bq_{2},\bq_{3}) \nonumber\\
 & \quad  +\left(\frac{1}{2}a_{\gamma a}^{(3)}(z)+a_{\gamma b}^{(3)}(z)\right)O_{\gamma\gamma}(\bq_{1},\bq_{2},\bq_{3})+ {\rm 3\;cyclic}\,   \label{eq:F3},
\end{align}
Those for the velocity divergence have the same structure,
\begin{align}
& 2\,G_{2}(\bq_1,\bq_2; z)   =2\, \beta(\bq_{1},\bq_{2})+ d_{\gamma}^{(2)}(z)\,\gamma(\bq_{1},\bq_{2})\,,\\
& 3!\, G_{3}(\bq_1,\bq_2,\bq_3; z)  =2\, O_{\beta\beta}(\bq_{1},\bq_{2},\bq_{3})
+\bigg[2d_{\gamma}^{(2)}(z)+d_{\gamma a}^{(3)}(z)\nonumber \\&\quad -2\left(d_{\gamma b}^{(3)}(z)+a_{\gamma}^{(2)}(z) - 1\right)\bigg] O_{\beta\gamma}(\bq_{1},\bq_{2},\bq_{3}) +\left(\frac{1}{4}d_{\gamma a}^{(3)}(z)-\frac{1}{2}d_{\gamma b}^{(3)}(z)\right)O_{\gamma\alpha_{a}}(\bq_{1},\bq_{2},\bq_{3})\nonumber\\
&\quad +\left(d_{\gamma b}^{(3)}(z)-\frac{1}{2}d_{\gamma a}^{(3)}(z)+2\left(a_{\gamma}^{(2)}(z) - 1\right)\right) O_{\gamma\beta}(\bq_{1},\bq_{2},\bq_{3})\nonumber\\
&\quad +\left(\frac{1}{2}d_{\gamma a}^{(3)}(z)+d_{\gamma b}^{(3)}(z)\right)O_{\gamma\gamma}(\bq_{1},\bq_{2},\bq_{3})+ {\rm 3\;cyclic}\,\label{eq:G3},
\end{align}
where
\begin{equation}
O_{XY}(\bq_{1},\bq_{2},\bq_{3})=X(\bq_{1},\bq_{2})Y(\bq_{12},\bq_{3})
\end{equation}
(and $\bq_{12}=\bq_{1}+\bq_{2}$).

To derive the mapping, we consider the redshift-space version of the relation between density and displacement field, 
\begin{equation}
    \delta^{S}(\bk,\tau) = \int \, d^3\bq e^{i \bk\cdot \bq}e^{i \bk\cdot \bm{\psi}^{S}(\bq,\tau)} - (2\pi)^3\delta_D(\bk)\,,
\end{equation}
and expand the exponential inside the integral to keep terms up to third order in the linear field, using the results obtained in Sects.~\ref{sect:PT} and~\ref{sect:RSD}.
The displacement field in redshift space, $\bm{\psi}^{S}(\bq,\tau) = \bm{\psi}(\bq, \tau) + \Delta_S\bm{\psi}(\bq, \tau)$, at each order is given by
\begin{align}
    \Delta_{S}\bm{\psi}^{(1)}_i(\bq, \tau) &= D(\tau)f(\tau)\hat{z}_i\hat{z}_j \phi_{,j}^{(1)}(\bq)\nonumber\\
    \Delta_{S}\bm{\psi}^{(2)}_i(\bq, \tau) &= D^2(\tau) f(\tau)c^{(2)}(\tau)\left(2 + \frac{d\log{c^{(2)}}}{d\log{D}}\right)\hat{z}_i\hat{z}_j \phi_{,j}^{(2)}(\bq)\nonumber\\
    \Delta_{S}\bm{\psi}^{(3)}_i(\bq, \tau) &= D^3(\tau) f(\tau)\hat{z}_i\hat{z}_j\Bigg[c_1^{(3)}(\tau)\left(3 + \frac{d\log{c_1^{(3)}}}{d\log{D}}\right)\phi_{1,j}^{(3)}(\bq)\nonumber\\
     &\qquad\qquad\qquad\quad\,\,+ c_2^{(3)}(\tau)\left(3 + \frac{d\log{c_2^{(3)}}}{d\log{D}}\right)\phi_{2,j}^{(3)}(\bq) \nonumber\\
     &\qquad\qquad\qquad\quad\,\,+d^{(3)}(\tau)\left(3 + \frac{d\log{d^{(3)}}}{d\log{D}}\right)v^{(3)}_j(\bq)\Bigg]
\end{align}
At first order we recover the Kaiser formula for dark matter \cite{Kaiser:1987qv},
\begin{equation}
    \delta^{\rm S} (\bk, \tau) =\big(1 + f(\tau) \mu_k^2\big)\delta^{(1)}(\bk)\,,
\end{equation}
while the second perturbative order in real space gives 
\begin{equation}
a_\gamma^{(2)} = 1 - c^{(2)}\,,
\end{equation}
where $a_\gamma^{(2)}$ is the eulerian bootstrap coefficient of the $\gamma(\bq_1, \bq_2)$ operator in the second order matter kernel, see Eq.~\re{eq:F2},  and $c^{(2)}$ the second order one in lagrangian bootstrap, introduced in Eq.~\re{eq:phi2}. We will keep the time dependence implicit from now on to avoid clutter. The analogous eulerian coefficient for the velocity kernel, $d_\gamma^{(2)}$, can be determined considering the redshift space contribution, 
\begin{equation}
    d_\gamma^{(2)}  =  - c^{(2)}\left(2 + \frac{d\log{c^{(2)}}}{d\log{D}}\right)\,.
\end{equation}
At third order, we obtain analogously,\footnote{The relations between these parameters and those used in  ~\cite{DAmico:2021rdb} is $a_{\gamma\alpha}^{(3)} = a_{\gamma a}^{(3)}/4 - a_{\gamma b}^{(3)}/2$, $a_{\gamma\gamma}^{(3)} = a_{\gamma a}^{(3)}/2 + a_{\gamma b}^{(3)}$.}
\begin{align}
 a_{\gamma a}^{(3)} -a_\gamma^{(2)} &=\frac{1}{2}c_2^{(3)}-1\,,\nonumber\\
  a_{\gamma a}^{(3)}-\frac{1}{2}a_\gamma^{(2)} &= \frac{1}{3} c_1^{(3)}+\frac{1}{4} c_2^{(3)}-\frac{1}{6}\,,
\end{align}
for matter and
\begin{align}
    d_{\gamma a}^{(3)} &= \frac{c_2^{(3)}}{2}\left(3 + \frac{d\log{c_2^{(3)}}}{d\log{D}}\right) +  \frac{c^{(2)}}{2}\left(1 + \frac{d\log{c^{(2)}}}{d\log{D}}\right)\,,\nonumber\\
    d_{\gamma b}^{(3)}  &= \frac{c_1^{(3)}}{3}\left(3 + \frac{d\log{c_1^{(3)}}}{d\log{D}}\right) + \frac{c_2^{(3)}}{4}\left(3 + \frac{d\log{c_2^{(3)}}}{d\log{D}}\right) - \frac{c^{(2)}}{2}\left(1 + \frac{d\log{c^{(2)}}}{d\log{D}}\right)\,,
\end{align}
for the velocity coefficients.
The coefficient of the transverse vector part $d^{(3)}$, see Eq.~\eqref{eq:vort3} can be fixed from the redshift space contributions, to obtain 
\begin{equation}
    d^{(3)} \left(3 + \frac{d\log{d^{(3)}}}{d\log{D}}\right)  = -c^{(2)} +c^{(2)}\left( 2 + \frac{d\log{c^{(2)}}}{d\log{D}}\right) = a_\gamma^{(2)} - 1 - d_\gamma^{(2)}\,,
\end{equation}
which can be integrated to give
\begin{equation}
    d^{(3)}(\tau) = -\frac{1}{3} + \int^{\tau}d\tau'\, \mathcal{H}(\tau)f(\tau)\left(\frac{D(\tau')}{D(\tau)}\right)^3 \left(a_\gamma^{(2)}(\tau') - d_\gamma^{(2)}(\tau')\right)\,,
\end{equation}
which confirms what we already discussed in Sect.~\ref{sect:vorto}, namely, that the vorticity part of the displacement field  is entirely determined by lower order scalar coefficients.

\bibliographystyle{JHEP2015}
\bibliography{mybib,biblio,references,newbib}
\label{lastpage}

\end{document}